\documentclass[aps,prb,reprint,a4paper,floatfix,groupedaddress,amsmath,amssymb,amsfonts]{revtex4-1}
\pdfoutput=1
\usepackage{mathptmx}
\usepackage{textcomp}
\usepackage{helvet}
\usepackage[utf8]{inputenc} 
\usepackage{graphicx}
\usepackage{upgreek} 
\usepackage{xspace}
\usepackage{textcomp}
\usepackage[pdftex]{hyperref}
\hypersetup{
	pdftitle={Exciton dynamics in GaAs/(Al,Ga)As core-shell nanowires with shell-quantum dots},
	colorlinks,
	citecolor=blue
	}
\usepackage[rightcaption]{sidecap}
\usepackage{color,soul} 
\usepackage[
,textwidth=17.5cm
,textheight=23.5cm
,verbose
,dvips
]{geometry}

\usepackage[colorinlistoftodos,prependcaption,textsize=footnotesize,textwidth=1.5cm]{todonotes}

\begin{document}
\title{Exciton dynamics in GaAs/(Al,Ga)As core-shell nanowires with shell quantum dots}

\author{Pierre Corfdir}
\email{corfdir@pdi-berlin.de}
\author{Hanno Küpers}
\author{Ryan B. Lewis}
\author{Timur Flissikowski}
\author{Holger T. Grahn}
\author{Lutz Geelhaar}
\author{Oliver Brandt}

\affiliation{Paul-Drude-Institut für Festkörperelektronik,
Hausvogteiplatz 5--7, 10117 Berlin, Germany}


\begin{abstract}

We study the dynamics of excitons in GaAs/(Al,Ga)As core-shell nanowires by continuous-wave and time-resolved photoluminescence and photoluminescence excitation spectroscopy. Strong Al segregation in the shell of the nanowires leads to the formation of Ga-rich inclusions acting as quantum dots. At 10~K, intense light emission associated with these shell quantum dots is observed. The average radiative lifetime of excitons confined in the shell quantum dots is 1.7~ns. We show that excitons may tunnel toward adjacent shell quantum dots and nonradiative point defects. We investigate the changes in the dynamics of charge carriers in the shell with increasing temperature, with particular emphasis on the transfer of carriers from the shell to the core of the nanowires. We finally discuss the implications of carrier localization in the (Al,Ga)As shell for fundamental studies and optoelectronic applications based on core-shell III-As nanowires.

\end{abstract}

\maketitle

\section{Introduction}

The relaxation and recombination dynamics of charge carriers in semiconductor quantum dots (QDs) has attracted much attention \cite{Ohnesorge1996,Merano2005} due to its consequences on the radiative efficiency of these zero-dimensional systems, which are used in quantum dot lasers\cite{Ledentsov1996} or single-photon and entangled-photon pair emitters.\cite{Michler2000,Stevenson2006} In particular, the interplay between the radiative decay of the exciton and the capture and escape of carriers from the QD controls the intensity, the lifetime, and the linewidth of the optical transitions related to the QD.\cite{Brusaferri1996,Findeis2001,Smith2005,VanHattem2013} In most cases, QDs form the lowest energy states of a given system. \cite{Ohnesorge1996,Warburton2000,Merano2005} Charge carriers excited nonresonantly in a continuum of states (e.g., the wetting layer for Stranski-Krastanov QDs) relax via phonon emission toward the QD, where they recombine radiatively. In striking contrast, there have been recent reports on new kinds of QD states lying hundreds of meV above the minimum of a continuum of states and yet giving rise to intense light emission at low temperatures.\cite{Heiss2013,Rudolph2013a,Fontana2014,Jeon2015} These QDs consist of Ga-rich inclusions that form spontaneously in the shells of GaAs/(Al,Ga)As core-shell nanowires, possibly as a result of differences in chemical potential and adatom mobilities between the \{111\} and \{112\} facets of the nanowires.\cite{Heiss2013,Mancini2014} In addition to the fact that they are bright single photon emitters, these so-called shell QDs are also suited for nano-sensing and opto-mechanical applications as they are located close to the surface and off-axis of the nanowires.\cite{Heiss2013,Montinaro2014} For all these applications, however, the shell QDs need to be populated by an exciton. For this purpose, one requires detailed knowledge on the dynamics of shell QD excitons and, in particular, on how efficient the radiative decay of shell QD excitons is as compared to their transfer to the GaAs core of the nanowires.

In this paper, we investigate the relaxation and recombination dynamics of charge carriers in core-shell GaAs/(Al,Ga)As nanowires by continuous-wave and time-resolved photoluminescence (PL) and photoluminescence excitation (PLE) spectroscopy. At 10~K, shell QDs can bind excitons and biexcitons that give rise to intense transitions in the 1.7~eV range. With increasing temperature, we observe a drastic quenching of the emission from the shell QDs. This quenching arises from the thermal activation of two nonradiative channels, which are associated with recombination at point defects in the shells and transfer of carriers from the shell to the core of the nanowire. We determine the characteristic times for the radiative and nonradiative decay channels of excitons in shell QDs, and we analyze their evolution with temperature.

\section{Experimental details}

\begin{figure}
\includegraphics[scale=1]{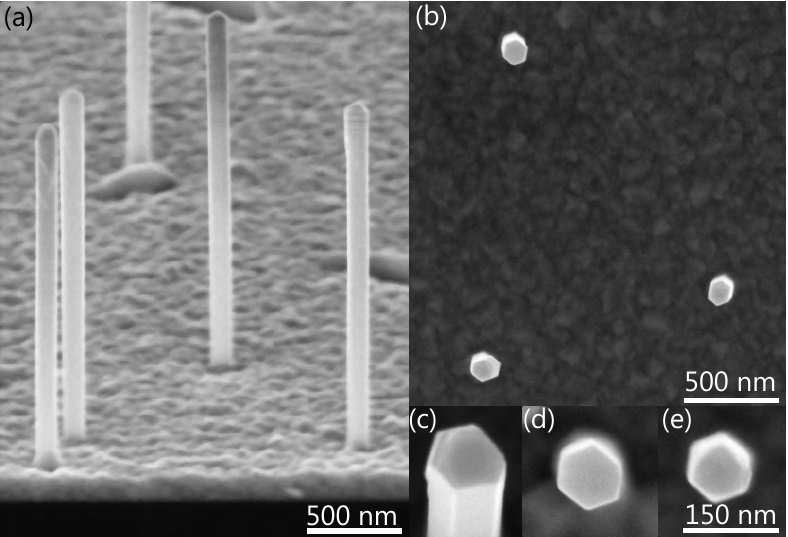}
\caption{Bird's eye view (a,c) and top-view (b,d,e) scanning electron micrographs from the GaAs/(Al,Ga)As core-shell nanowires studied in this work.}
\label{fig:FigureSEM}
\end{figure}

The nanowires were grown by solid-source molecular beam epitaxy on an n-type Si(111) substrate. Prior to growth, the substrate was annealed in the growth chamber for 10~min at 650\,°C. The substrate temperature was then decreased to 630\,°C, after which Ga was deposited for 30~s at a GaAs equivalent growth rate of 0.44~monolayer/s of Ga in the lattice structure of GaAs, and the substrate was subsequently annealed for 60~s to form droplets. The growth of GaAs nanowires was initiated by supplying Ga and As$_2$ for 50~min with fluxes of 0.11 and 0.75~monolayer/s, respectively. To stop the axial growth, the Ga flux was interrupted, and the Ga droplet was consumed at a low As$_2$ flux, while the substrate temperature was decreased to 500\,°C. Finally, a 30~nm thick Al$_{0.25}$Ga$_{0.75}$As shell was grown using a combined group-III flux of 0.6~monolayer/s and a V/III ratio of 5. The morphology of the nanowires was investigated with a field emission gun scanning electron microscope using an acceleration voltage of 5 kV. As depicted in Fig.~\ref{fig:FigureSEM}, the average length and diameter of our nanowires are 2~\textmu m and 120~nm, respectively. The areal density in nanowires is about 0.1~\textmu m$^{-2}$.

Continuous-wave PL experiments were performed using a HeNe laser emitting at 632.8~nm. For experiments on nanowire ensembles, the laser was focused onto the sample using a plano-convex lens with a focal length of 5~cm, and the excitation power was varied between 0.005 and 6.6~mW. For single nanowire experiments, the laser light was focused using a microscope objective with a numerical aperture of 0.7, and the excitation power was varied between 57~nW and 8.1~mW. The PL signal was dispersed using a monochromator with a spectral resolution of 0.5~meV and detected by a charge-coupled device camera. Tunable excitation for PLE experiments was obtained from a 100~W halogen lamp dispersed by a monochromator with a 64~cm focal length. Time-resolved PL spectroscopy was carried out using fs pulses obtained from an optical parametric oscillator pumped by a Ti:sapphire laser. The emission wavelength and repetition rate were 632.8~nm and 76~MHz, respectively, and the energy fluence per pulse was kept below 0.7~\textmu J /cm$^2$. The transient emission was spectrally dispersed by a monochromator and detected by a streak camera operating in synchroscan mode. In all PL and PLE measurements, the samples were mounted in a coldfinger cryostat, whose temperature can be varied between 10 and 300~K. Approximately 100 nanowires were probed in experiments on ensembles of nanowires.

\section{Results and discussion}

\begin{figure}
\includegraphics[scale=1]{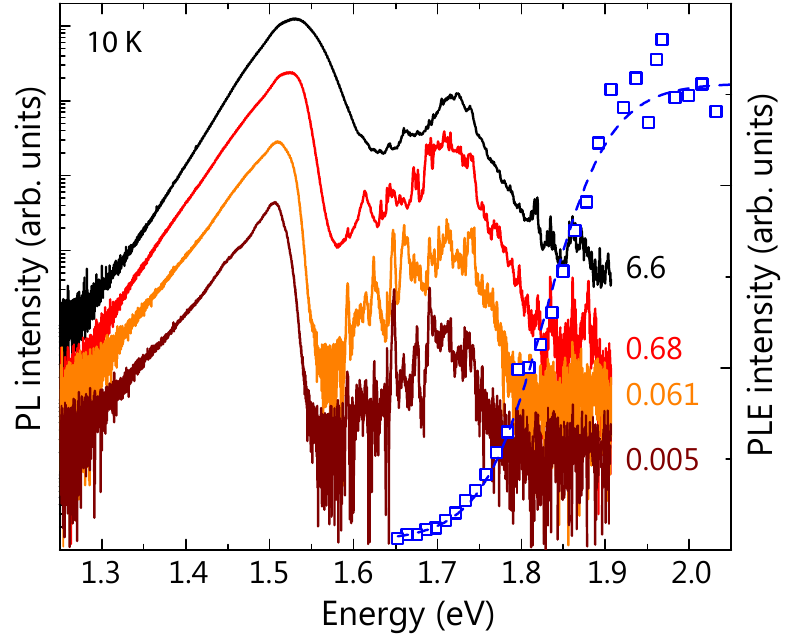}
\caption{PL spectra acquired at 10~K at various excitation powers from an ensemble of GaAs/(Al,Ga)As core-shell nanowires (solid lines). The numbers on the right indicate the excitation power in mW. The empty squares show a PLE spectrum detected at 1.505~eV and measured at 10~K. The dotted line is a fit to the PLE spectrum using a logistic function, yielding an (Al,Ga)As shells bandgap and an Urbach tail energy of 1.83~eV and $(38 \pm 3)$~meV, respectively.}
\label{fig:Figure1}
\end{figure}

Figure~\ref{fig:Figure1} shows a typical PL spectrum of an ensemble of GaAs/(Al,Ga)As core-shell nanowires at 10~K for excitation powers between 0.005 and 6.6~mW. Two distinct bands centered around 1.5 and 1.7~eV are detected. We attribute the intense band at about 1.5~eV to carrier recombination in the GaAs core. For an excitation power of 0.005~mW, the emission from the core is centered at 1.505~eV and arises from recombination within polytypic nanowire segments.\cite{Graham2013} For larger excitation powers, this emission band blueshifts, as a result of filling of the states related to the polytypic segments, and is centered at 1.529~eV for the highest excitation power. The latter energy is significantly larger than the bandgap of strain-free zincblende GaAs (1.519~eV). Although not understood yet, this behavior has been commonly reported for GaAs nanowires (see Ref.~\onlinecite{Martelli2015} for a detailed discussion). At high excitation powers, the band on the high-energy side of the PL line from the GaAs core is centered at 1.717~eV and exhibits a linewidth of 53~meV. Decreasing the excitation power reveals that this band is actually composed of several tens of narrow transitions with a typical linewidth on the order of 1~meV, arising from the recombination of charge carriers confined at localized states that we attribute to  Ga-rich inclusions that form spontaneously in the (Al,Ga)As shells of the nanowires.\citep{Heiss2013,Rudolph2013a,Fontana2014} In agreement with the results of Refs.~\onlinecite{Heiss2013,Fontana2014,Rudolph2013a,Jeon2015,Ramsteiner1997}, we attribute these localized states to in the (Al,Ga)As shells.

\begin{SCfigure*}
\includegraphics[scale=1]{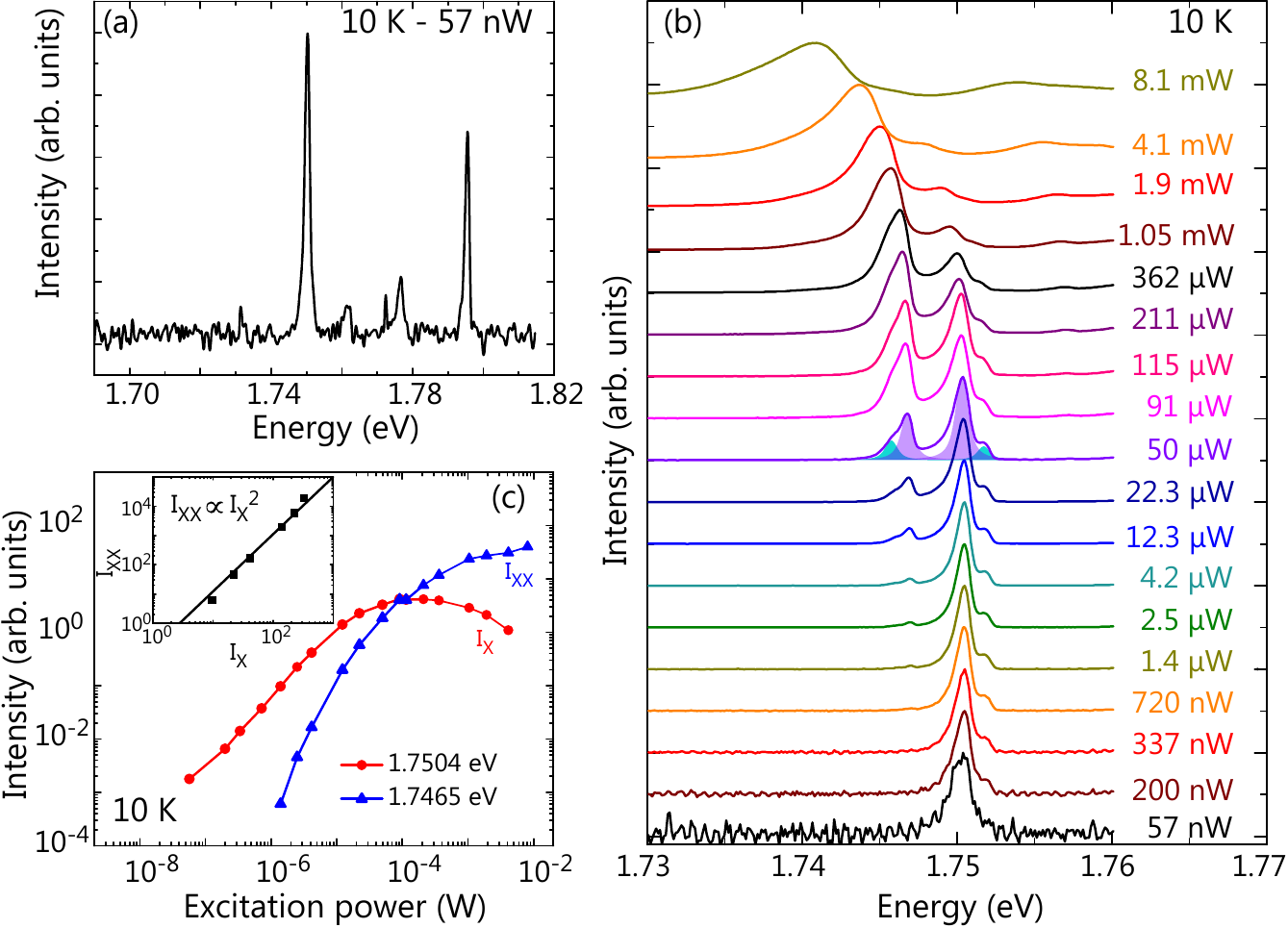}
\caption{(a) PL spectrum from a single nanowire taken at 10~K with an excitation power of 57~nW. (b) Evolution of the single nanowire PL spectra with excitation power between 57~nW and 8.1~mW. The blue and violet shaded areas highlight the fact that the peaks at 1.7465 and 1.7504~eV are actually doublets. (c) Dependence of the intensity of the transitions at 1.7465 and 1.7504~eV observed in (b) on excitation power. Inset: PL intensity at 1.7465~eV ($I_\text{XX}$) as a function of that of the line at 1.7504~eV ($I_\text{X}$). The solid line highlights the quadratic dependence of $I_\text{XX}$ on $I_\text{X}$.}
\label{fig:Figure2}
\end{SCfigure*}

To gain insight into the electronic properties of these so-called shell QDs, we have performed single nanowire spectroscopy. Figure~\ref{fig:Figure2}(a) shows the (Al,Ga)As shell luminescence from a single nanowire for an excitation power of 57~nW. When excited at this low excitation power, this nanowire exhibits only a few transitions between 1.7 and 1.8~eV. The linewidth of these individual lines is on the order of 1~meV, which is orders of magnitude larger than the value expected for the radiative linewidth of an exciton in a shell QD.\citep{Fontana2014} This linewidth is also significantly larger than that reported for the shell QDs in Refs.~\onlinecite{Rudolph2013a,Fontana2014}. As shown in Refs.~\onlinecite{Weiss2014,Francaviglia2015}, shell QDs in the first 20~nm of the shells are optically inactive. Considering the fact that the shell thickness in our nanowires is 30~nm and that our nanowires have been grown without GaAs passivation shell, the shell QDs giving rise to the transitions observed in Figs.~\ref{fig:Figure1} and \ref{fig:Figure2}(a) are most likely located close to the surface. The actual confining potential in the shell QDs is thus not only given by the spatial distribution of Al in the shells, but also by the surface potential arising from the Fermi level pinning at the nanowire sidewalls,\citep{Corfdir2014} and we tentatively attribute the large emission linewidth of our shell QDs to spectral diffusion due to electrostatic fluctuations arising from photo-induced changes in the surface states of the nanowires.\cite{Holmes2015} A statistical analysis of the PL spectra of 100 nanowires (not shown) reveals that each nanowire contains on average 13 optically active shell QDs per nanowire. However, as QDs located in the innermost 20~nm of the shells\cite{Francaviglia2015} and at the surface do not contribute to the PL spectrum, it is likely that the total number of shell QDs in our nanowires is much larger.

Figure~\ref{fig:Figure2}(b) shows the evolution with excitation power of the PL at 10~K from a shell QD emitting at 1.7504~eV. The evolution of the intensity with excitation power of the transitions related to this QD is shown in Fig.~\ref{fig:Figure2}(c). For excitation powers not exceeding 22~\textmu W, the PL intensity of the transition at 1.7504~eV increases nearly linearly with excitation power [Fig.~\ref{fig:Figure2}(c)], indicating that this line arises from the recombination of a neutral exciton bound to a shell QD. For higher excitation powers, the intensity of the exciton saturates and eventually decreases. For excitation powers larger than 1.4~\textmu W, one detects an additional line at 1.7469~eV whose intensity increases quadratically with the intensity of the neutral exciton, demonstrating its biexcitonic nature [Figs.~\ref{fig:Figure2}(b) and \ref{fig:Figure2}(c)]. At high excitation powers, shell QDs after recombination of the biexciton can be populated with another electron-hole pair prior to the recombination of the exciton, leading to the decrease in exciton intensity for powers larger than 211~\textmu W. Note that heating effects at high excitation powers (see the redhshift of the exciton transition in Fig.~\ref{fig:Figure2}(c)] can also result in the observed decrease in the intensity of the exciton. We deduce that the biexciton binding energy in our shell QDs is 3.5~meV, a value significantly lower than the 6~meV reported for (Al,Ga)As shells with an average Al content of 51\%.\cite{Fontana2014} Since the biexciton binding energy increases with localization depth,\cite{Langbein1999} we ascribe the weaker biexciton binding in our shell QDs to the lower average Al content in the shells of our nanowires (25\%).

As highlighted for the spectrum taken with an excitation power of 50~\textmu W in Fig.~\ref{fig:Figure2}(b), the exciton and biexciton transitions exhibit a shoulder on their high and low energy sides, respectively. In both cases, the energy difference and the intensity ratio between the main peak and its shoulder are 1~meV and about 5, respectively. We suggest that the splittings observed for the exciton and biexciton transitions arise from the anisotropic exchange splitting of the bright states of the shell QD exciton.\cite{Seguin2005} Note that the asymmetry in the intensity of the exciton and biexciton doublets is a result of the antenna effect, as already discussed in Ref.~\onlinecite{Fontana2014}. While a 1~meV splitting between the bright states of the exciton is larger than previous values reported for shell QDs,\cite{Fontana2014} similar splittings have already been reported for group-III-As based QDs.\cite{Finley2002} This large anisotropic exchange splitting confirms the results of Refs.~\onlinecite{Corfdir2014,Jeon2015} that shell-QDs are small and elongated. 

It is worth emphasizing that the QD-like transitions in Fig.~\ref{fig:Figure2} have been detected at energies corresponding to the high-energy side of the PL band of the (Al,Ga)As shell in a nanowire ensemble. Therefore, both the low and high energy tails of the (Al,Ga)As PL band in Fig.~\ref{fig:Figure1} are related to deeply localized states. The PLE spectrum shown in Fig.~\ref{fig:Figure1} and measured at 1.505~eV on an ensemble of nanowires confirms this finding. For excitation energies above 1.8~eV, the PL intensity from the GaAs core increases significantly, due to absorption of carriers by the shells. Fitting the PLE spectrum with a logistic function\cite{ODonnell1999} yields an (Al,Ga)As shells bandgap and an Urbach energy of 1.83~eV and $(38 \pm 3)$~meV, respectively. Since the PL from the (Al,Ga)As shell is centered at 1.717~eV, the low and high energy tails of the (Al,Ga)As PL band both originate from the recombination of excitons deeply localized in shell QDs.

Despite the fact that the discrete electronic states associated with the shell QDs lie approximately 200~meV above the continuum of states related to the GaAs core, they manifest themselves in an intense PL signal (see Fig.~\ref{fig:Figure1} as well as Refs.~\onlinecite{Heiss2013,Fontana2014}). For instance, the PL intensity ratio between the shell QDs and the GaAs core for the ensemble of nanowires is equal to $9 \times 10^{-2}$ at 10 K (Fig.~\ref{fig:Figure2}). At thermal equilibrium, this ratio should take a much smaller value given by $\exp \left( - \Delta E /kT \right)$, where $\Delta E$ is the difference between the PL peak energies of the shell QDs and the GaAs core. Even if the combination between heavy-hole / light-hole mixing in shell QDs\cite{Fontana2014} and the nanowire antenna effect may result in a light excitation efficiency larger for the shell QDs than for the GaAs core,\cite{Corfdir2015} the strong PL from the shell QDs mainly originates from the fact that the transfer of excitons from the shell to the GaAs core is hindered by their localization at shell QDs.
 
\begin{figure*}
\includegraphics[scale=1]{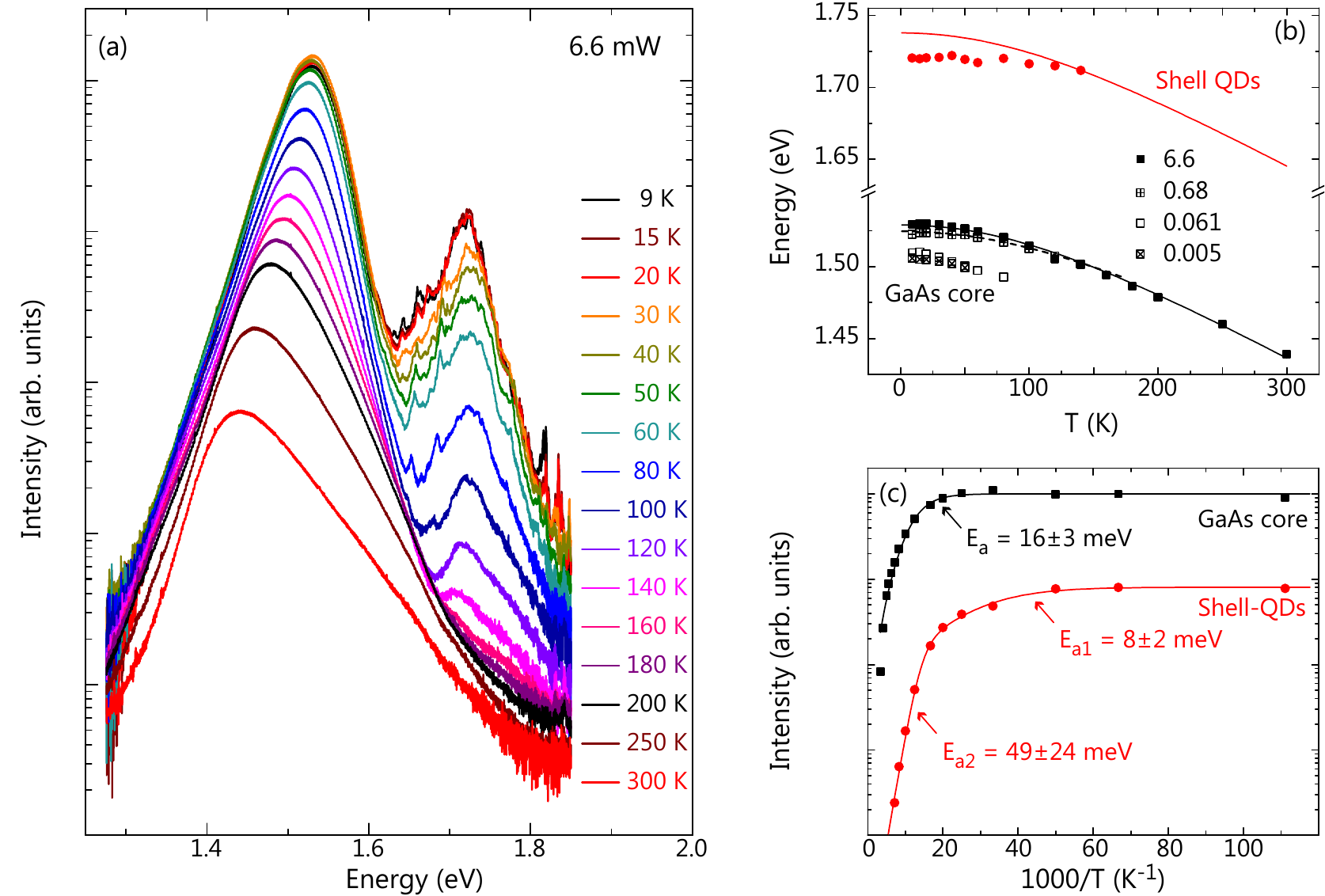}
\caption{(a) PL spectra as a function of temperature for an ensemble of GaAs/(Al,Ga)As core-shell nanowires taken with an excitation power of 6.6~mW. (b) Temperature dependence of the peak energy of the PL band from the GaAs core (squares) and the (Al,Ga)As shells (circles). The numbers on the right indicate the excitation power in mW. The solid and dotted lines show the results of a fit using Eq.~(14) in Ref.~\onlinecite{Passler1997} for the energies measured with an excitation power of 6.6 and 0.68~mW, respectively. For the excitation power of 6.6~mW, the fit yields $\upalpha = 0.51$~meV/K and $\uptheta=296$~K. (c) Arrhenius plot showing the temperature dependence of the PL band from the GaAs core (squares) and the (Al,Ga)As shells (circles) taken with an excitation power of 6.6~mW. The solid and dashed lines are fits using Eqs.~\ref{eq:IPL1} and \ref{eq:IPL2}, respectively.}
\label{fig:Figure3}
\end{figure*}

For obtaining a detailed understanding of the transfer dynamics of carriers in GaAs/(Al,Ga)As core-shell nanowires, we have performed temperature-dependent experiments. Figure~\ref{fig:Figure3}(a) shows the temperature dependence of the PL from a nanowire ensemble for an excitation power of 6.6~mW. As displayed in Fig.~\ref{fig:Figure3}(b), the PL energy of the GaAs core redshifts monotonically with increasing temperature. Its temperature dependence can be described well by the expressions derived by \citet{Passler1997} with the parameters recommended in Ref.~\onlinecite{Passler1999} for bulk GaAs. This finding, however, is most probably fortuitous. As shown in Fig.~\ref{fig:Figure3}(a), a different behavior is observed when using an excitation power of 0.68~mW. The temperature dependence of the PL energy of GaAs grown in the form of nanowires therefore depends not only on the lattice structure\cite{Zilli2015} and on the detailed growth conditions,\cite{Martelli2015} but also on the excitation power. The energy of the transitions associated with the shell QDs remains nearly constant between 9 and 100~K. This deviation from the temperature dependence of the gap has been commonly reported for disordered systems\cite{Cho1998,Graham2013} and confirms that in this temperature range transfer processes of excitons occur within a band of localized states in the (Al,Ga)As shell. 

\begin{figure*}
\includegraphics[width=\textwidth,height=\textheight,keepaspectratio]{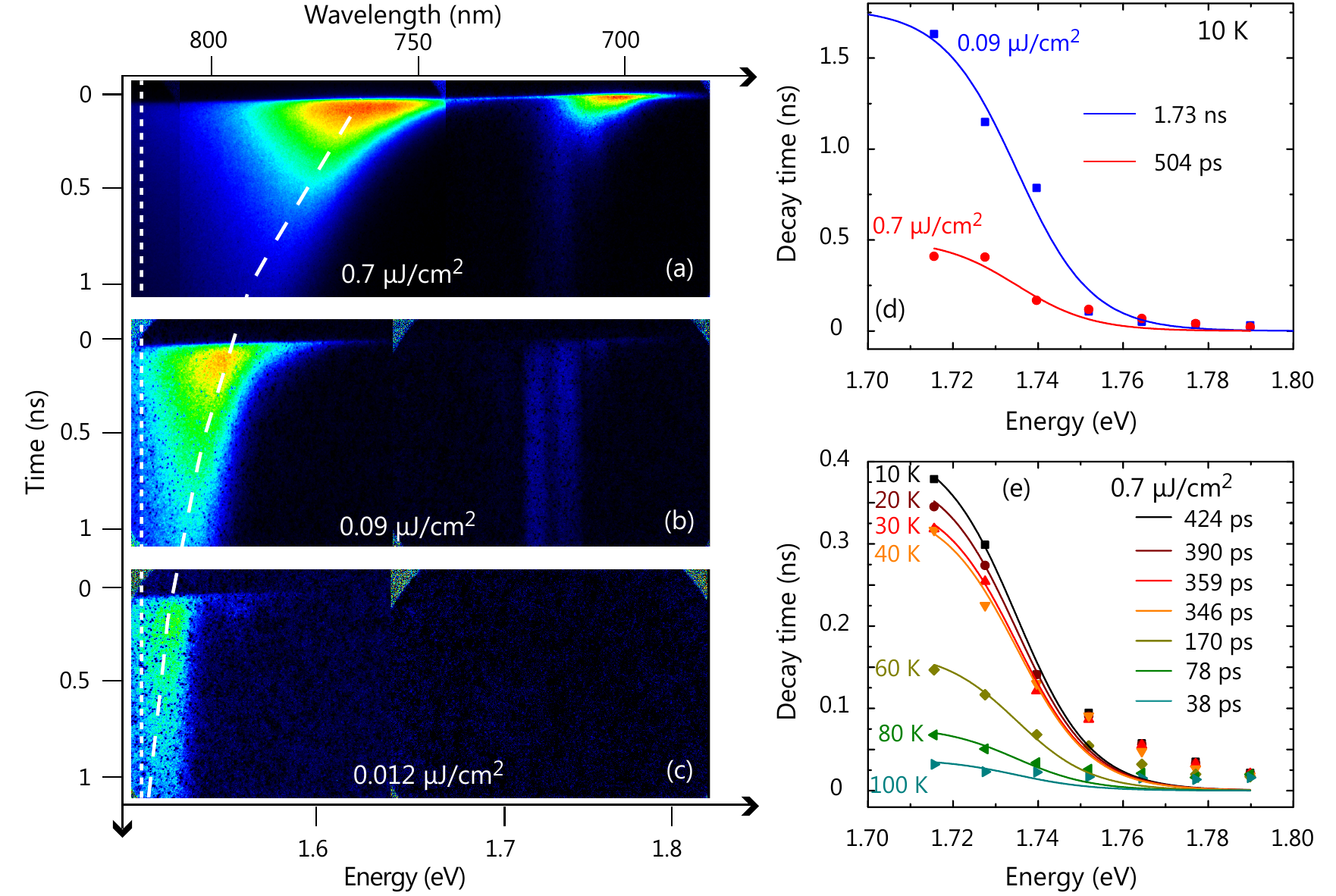}
\caption{Streak camera images of an ensemble of nanowires acquired at 10~K with an energy fluence per pulse of (a) 0.7, (b) 0.09, and (c) 0.012~\textmu J/cm$^2$ (c). The intensity is color-coded from black (low intensity) to red (high intensity). The vertical dotted line shows the value of the bandgap of strain-free zincblende GaAs, and the dashed line highlights the shift of the GaAs core PL with increasing time and excitation power. (d) Energy dependence of the PL decay time of the shell QDs for an energy fluence per pulse of 0.09 and 0.7~\textmu J/cm$^2$ (squares and circles, respectively). (e) Energy dependence of the PL decay time of the shell QDs for various temperatures and an energy fluence per pulse of 0.7~\textmu J/cm$^2$. The solid lines in (d) and (e) are the results of fits using Eq.~(\ref{eq:Gourdon}) with $E_\text{m}=1.735$~eV and $E_0=9$~meV, and the numbers on the right indicate $\tau_\text{QD}$.}
\label{fig:Figure4}
\end{figure*}

Figure~\ref{fig:Figure3}(c) shows the evolution of the integrated PL intensity as a function of temperature of the GaAs core and of the shell QDs. The former is almost constant between 10 and 50~K and decreases for higher temperatures. This quenching arises from the combined decrease in the radiative recombination rate of carriers in the core and from the activation of a nonradiative recombination channel with activation energy $E_\text{A}$. Since the core diameter is 60~nm, the density of states for carriers in the GaAs core is three-dimensional. The radiative recombination rate of carriers is thus proportional to $T^{-3/2}$, and the temperature dependence of emission intensity of the GaAs core ($I_\text{c}$) can  be written as:\cite{Hauswald2014}

\begin{equation}\label{eq:IPL1}
I_\text{c} \propto \left(1+aT^{3/2} e^{- E_\text{A} / k_B T} \right)^{-1},
\end{equation}

\noindent
where $a$ is a fit parameter. As shown in Fig.~\ref{fig:Figure3}(c), the data are fit well up to 200~K with $E_\text{A} = (16 \pm 3)$~meV. This value coincides with those reported in Refs.~\onlinecite{Titova2006,Rudolph2013b} for GaAs/(Al,Ga)As core-shell nanowires, despite different fabrication routes and different nanowire lengths, shell thicknesses, and Al contents. The fact that the PL intensity of all of these samples quenches with the same activation energy strongly suggests a common microscopic origin for the nonradiative process in these structures. Considering the small diameter of the GaAs core, we believe it is likely that a specific recombination center at the GaAs/(Al,Ga)As \{110\} interfaces controls this process.

The PL intensity of the shell QDs ($I_\text{QD}$) also decreases with increasing temperature. However, the overall quenching with temperature is much stronger for $I_\text{QD}$ than for $I_\text{c}$: the ratio $I_\text{QD}/I_\text{c}$ decreases from $9 \times 10^{-2}$ to $1.5 \times 10^{-3}$ when the temperature increases from 9 to 140~K. This behavior is similar to what has been reported for planar quantum wells with disordered barriers:\cite{,Ramsteiner1997,Corfdir2012} while transitions related to the quantum well and the barriers are detected at low temperatures, an increase in temperature leads to an increase in the efficiency in carrier capture and to a fast quenching of the barrier luminescence intensity. Accordingly, we ascribe the strong decrease in $I_\text{QD}/I_\text{c}$ to an increasingly efficient carrier transfer from the shell to the core with increasing temperature.

While the decrease in $I_\text{QD}$ is weak between 10 and 50~K, it is much stronger at higher temperatures. Two thermally activated nonradiative recombination channels need to be taken into account to reproduce this temperature dependence, and we propose the two following phenomena for being responsible for the quenching of $I_\text{QD}$. First, a carrier in a shell QD may reach an adjacent QD by phonon-assisted tunneling. If this second QD is dark, i.e. it is located close enough to the surface, to the core or to a point defect, this process with an activation energy $E_\text{A1}$ gives rise to a decrease in the intensity of the shell QD PL band. Second, carriers can escape from the shell QDs to the (Al,Ga)As shells' continuum of states, a mechanism with an activation energy E$_{A2}$, and then be trapped by the GaAs core or the surface. We expect that $E_{A1} < E_{A2}$, since the former energy is given by the energy difference between two adjacent shell QDs and the latter by the energy difference between a shell QD and the bottom of the (Al,Ga)As shell continuum of states. Note that the radiative lifetime of the shell QD may also increase with temperature due to occupation of higher energy states that are dark,\cite{Yu1996,Suffczynski2009} contributing to the decrease in $I_\text{QD}$. This increase in radiative lifetime is however moderate,\cite{Yu1996} and we assume for simplicity that it is in fact constant with temperature. The evolution of $I_\text{QD}$ is thus given by:

\begin{equation}\label{eq:IPL2}
I_\text{QD} \propto \left(1+a_1 e^{-E_\text{A1} / kT}+ a_2 e^{-E_\text{A2} / kT} \right)^{-1},
\end{equation}

\noindent
where $a_1$ and $a_2$ are fitting parameters. The best fit to the data yields $E_\text{A1} = (8 \pm 2)$~meV and $E_\text{A2} = (49 \pm 24)$~meV. At cryogenic temperatures, carriers are efficiently trapped by shell QDs. Transfer between adjacent shell QDs may occur by tunneling, provided they are spatially and energetically close enough. In contrast, above 50~K, carriers have enough thermal energy to escape the QDs towards the shell continuum of states and be transferred to the GaAs core where they recombine.

The continuous-wave experiments in Fig.~\ref{fig:Figure3} only provide qualitative knowledge on the dynamics of carriers in shell QDs. Time-resolved PL experiments are needed to obtain quantitative information on the radiative lifetime of excitons in shell QDs and on the efficiency of the carrier transfer from the (Al,Ga)As shell to the GaAs core of the nanowires. Figures~\ref{fig:Figure4}(a)--\ref{fig:Figure4}(c) display time-resolved PL streak camera images at 10~K on an ensemble of nanowires for various excitation densities [energy fluence per pulse of 0.7, 0.09 and 0.012~\textmu J/cm$^2$ in Figs.~\ref{fig:Figure4}(a), \ref{fig:Figure4}(b) and \ref{fig:Figure4}(c), respectively]. An exponential fit to the spectrally integrated PL from the GaAs core after long delays yields a decay time of about 800~ps. This decay time corresponds to a recombination velocity of 1900~cm/s at the GaAs/(Al,Ga)As\{110\} interface. This value is more than an order of magnitude larger than the recombination velocities achieved for state-of-the-art GaAs/(Al,Ga)As\{100\} interfaces,\cite{Wolford1994} but similar to the one previously reported for GaAs/(Al,Ga)As core-shell nanowires.\cite{Demichel2010} This result is thus consistent with the hypothesis that the nonradiative process observed in Fig.~\ref{fig:Figure3}(c) for the GaAs core is governed by interface recombination.

The PL bands at zero delay of both the GaAs core and the shell QDs strongly blueshift with increasing excitation density. Fitting with an exponential the high-energy tail of the GaAs PL spectrum at zero time delay taken with an energy fluence per pulse of $0.7$~\textmu J/cm$^2$, we find that the effective carrier temperature at the early stage of the decay is about 300~K. The PL spectrum of the GaAs core at zero time delay is thus dominated by hot carrier luminescence.\cite{Xu1984,Pelouch1992} With increasing time delay, the GaAs core transition redshifts, and its high-energy tail gets steeper, as a result of the cooling of carriers.

To verify that the significant blueshift with increasing excitation of the PL spectra at zero time delay in Figs.~\ref{fig:Figure4}(a)-\ref{fig:Figure4}(c) is due to band filling in the GaAs core, we need to determine the density of photogenerated carriers. Such a determination is usually achieved based on an analysis of the PL spectra (see, for instance, Refs.~\onlinecite{Szczytko2004,Kappei2005,Amo2007}) rather than on simple estimates using the energy fluence per pulse. However, the direct applications of these methods is hindered by the large broadening of the PL from the GaAs core of our nanowires. Therefore, we have measured a 1~\textmu m thick GaAs epilayer sandwiched between AlAs using the same excitation conditions to accurately determine the photogenerated carrier density. To improve accuracy, the latter was obtained using two independent methods. While the first one is based on a lineshape fit of the PL spectra at zero delay, assuming parabolic conduction and valence bands and considering only k-conserving transitions,\cite{Kappei2005} the second relies on the analysis of the rise of the PL transient.\cite{Amo2007} Note that these methods are accurate for large excitation densities. In particular, they yield that excitation at 632.8~nm with an energy fluence per pulse of 70~\textmu J/cm$^2$ corresponds to a photogenerated carrier density of $2.0 \times 10^{16}$~cm$^{-3}$. Therefore, we deduce that excitation with an energy fluence per pulse of 0.7, 0.09 and 0.012~\textmu J/cm$^2$ should give rise to a photogenerated carrier density of $2.0 \times 10^{14}$, $2.6 \times 10^{13}$ and $3.4 \times 10^{12}$~cm$^{-3}$, respectively.\footnote{The measured carrier densities are significantly smaller than those obtained from simple estimates based on the energy fluence per pulse. These differences arise from the reflection of the laser from the surface, surface and interface recombination, diffusion and plasma expansion.} As shown in Ref.~\onlinecite{Heiss2014}, a nanowire with a cross-section $S$ exhibits an enhanced light absorption compared to the bulk. This absorption enhancement is given by the ratio between the power absorbed by the nanowire and the product of the incident power surface density with $S$. The absorption enhancement for light with a wavelength of 632.8~nm and polarized perpendicular to a GaAs nanowire with a diameter of 120~nm is equal to 50.\cite{Heiss2014} Assuming that photogenerated carriers are homogeneously distributed along the nanowire, the injected carrier densities in the case of Figs.~\ref{fig:Figure4}(a), \ref{fig:Figure4}(b) and \ref{fig:Figure4}(c) should not exceed $1.0 \times 10^{16}$, $1.3 \times 10^{15}$, and $1.7 \times 10^{14}$~cm$^{-3}$, respectively. Even the first of these densities is rather moderate, and the blueshift of about 100~meV in Fig.~\ref{fig:Figure4}(a) is thus rather surprising. Moreover, the cooling is seen to take place on unusually long time scales. For example, despite the very low initial carrier density corresponding to an energy fluence per pulse of 0.12~\textmu J/cm$^2$, it takes 1~ns for the emission from the GaAs core to reach the energy of strain-free zincblende GaAs [Fig.~\ref{fig:Figure4}(c)]. This finding suggests that the cooling of carriers in GaAs nanowires is drastically slower than in the bulk. The recent observation of long-lived hot carrier luminescence in GaAs and InP by \citet{Tedeschi2016} is consistent with this idea. The above bandgap emission observed in Fig.~\ref{fig:Figure1} and often reported for continuous-wave excitation in the literature (for an overview, see Ref.~\onlinecite{Martelli2015}) could thus simply arise from hot carrier luminescence. However, even when using a weak continuous-wave excitation, the luminescence of the GaAs core of our nanowires is broad (Fig.~\ref{fig:Figure1}). This inhomogeneous broadening impedes the extraction of an effective carrier temperature at long time delays and/or weak excitation conditions. Further experiments are thus required to ascertain whether the cooling rate of carriers is affected by the nanowire geometry.

\begin{SCfigure*}
\includegraphics[scale=1]{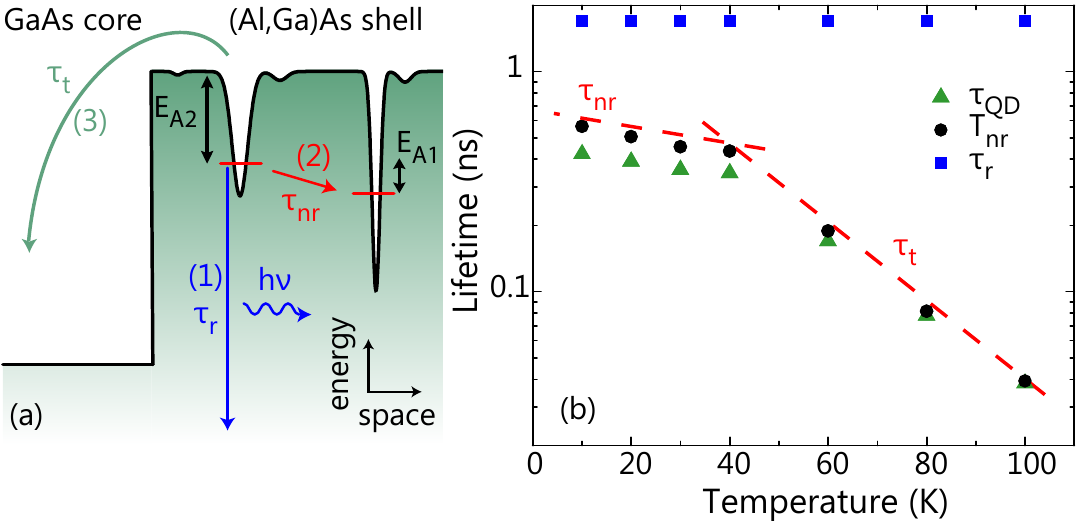}
\caption{(a) Recombination dynamics of excitons in the shell of GaAs/(Al,Ga)As nanowires. At low temperatures, excitons are localized at shell QDs. Radiative recombination (with a lifetime $\tau_\text{r}$), phonon-assisted QD-to-QD tunneling of excitons, and thermally activated transfer of carriers from the shell to the core of the nanowire (with characteristic time $\tau_\text{t}$ and activation energy $E_{A2}$) are shown by arrows (1), (2) and (3), respectively. If after tunneling the exciton is close to the surface or a point defect, process (2) is nonradiative, with lifetime and activation energy $\tau_{nr}$ and $E_{A1}$, respectively. (b) Temperature dependence of $\tau_\text{r}$, $T_\text{nr}$ and $\tau_\text{QD}$. The lines are a guide to the eye showing the temperature dependence of $\tau_\text{nr}$ and $\tau_\text{t}$.}
\label{fig:Figure5}
\end{SCfigure*} 

The energy dependence of the PL decay time $\tau_\text{PL}$ of the shell QDs is displayed in Fig.~\ref{fig:Figure4}(d) for an energy fluence per pulse of 0.09 and 0.7~\textmu J/cm$^2$. In both cases, $\tau_\text{PL}$ increases with decreasing energy. In particular, for an energy fluence per pulse of 0.09~\textmu J/cm$^2$, $\tau_\text{PL}$ is equal to 30~ps at 1.790~eV and increases to 1.63~ns at 1.715~eV. The behavior in Fig.~\ref{fig:Figure4}(d) is characteristic of the transfer of excitons between localized states,\citep{Gourdon1989} and it indicates that even at low temperatures exciton transfer between adjacent shell QDs is possible through tunneling assisted by acoustic phonons. Following the formalism described in Ref.~\onlinecite{Gourdon1989}, the energy dependence of $\tau_\text{PL}$ is given by

\begin{equation}\label{eq:Gourdon}
\frac{1}{\tau_\text{PL}} = \frac{1}{\tau_\text{QD}}\left(1 +e^{\left( E - E_\text{m} \right)/E_0}\right),
\end{equation}

\noindent
with $\tau_\text{QD}$ denoting the lifetime of the shell QD excitons in the absence of any tunneling between adjacent QDs, and $E_\text{m}$ the energy at which the tunneling time is equal to $\tau_\text{QD}$. The energy $E_0$ is characteristic of the density of states of the shell QDs and of the transfer probability of excitons between adjacent shell QDs. It is therefore independent of the localization depth of the shell QDs. It is also assumed in Eq.~\ref{eq:Gourdon} that $\tau_\text{QD}$ does not depend on the emission energy of the shell QD. This assumption is justified by the fact that the energy of the shell QDs measured in Fig.~\ref{fig:Figure4}(d) is well below the mobility edge of the (Al,Ga)As shell (Fig.~\ref{fig:Figure1}). The best fit to the data yields $\tau_\text{QD}=1.7$~ns, $E_\text{m}=1.735$~eV, and $E_0=9$~meV [Fig.~\ref{fig:Figure4}(d)]. The characteristic energy $E_0$ agrees well with the activation energy $E_\text{A1}$ measured for the initial quenching of the the emission from the shell QDs in Fig.~\ref{fig:Figure3}(c). This agreement confirms that this initial quenching is caused by a transfer of excitons within the shell followed by their recombination at point defects, rather than by an increase in the trapping efficiency of carriers by the core of the nanowires. Assuming that at 10~K and for an energy fluence per pulse of 0.09~\textmu J/cm$^2$ $\tau_\text{QD}$ is equal to the radiative lifetime $\tau_\text{r}$ of the shell QD exciton,  we obtain $\tau_\text{r} = 1.7$~ns, a value significantly larger than the value reported in Ref~\onlinecite{Heiss2013}.

As shown in Figs.~\ref{fig:Figure4}(d) and \ref{fig:Figure4}(e), $\tau_\text{PL}$ at a given energy decreases with increasing energy fluence per pulse and temperature, respectively. Since $E_\text{m}$ and $E_0$ are characteristic for the density of states of the shell QDs, we assume that they are independent of excitation density and temperature between 10 and 100~K, and the whole set of data in Figs.~\ref{fig:Figure4}(d) and \ref{fig:Figure4}(e) can be fit by only varying $\tau_\text{QD}$. The value of $\tau_\text{QD}$ decreases from 1.73 to 0.50~ns with increasing energy fluence, and from 424~ps to 38~ps with increasing temperature. We attribute both of these results to an increase in the efficiency of nonradiative recombination at point defects and in carrier transfer from the shell to the core with increasing excitation density and temperature. The observed decrease in nonradiative lifetime with increasing excitation density is probably a consequence of the slow cooling of carriers observed in Figs.~\ref{fig:Figure4}(a)--\ref{fig:Figure4}(c), as the latter results in an increase in effective temperature of carriers with increasing excitation.

Figure~\ref{fig:Figure5}(a) summarizes our understanding of the dynamics of excitons in core-shell GaAs/(Al,Ga)As nanowires. The lifetime $\tau_\text{QD}$ measured in Figs.~\ref{fig:Figure4}(d) and \ref{fig:Figure4}(e) depends on the radiative lifetime $\tau_\text{r}$, on the nonradiative lifetime at point defects $\tau_\text{nr}$, and on the characteristic transfer time $\tau_\text{t}$ of carriers from the shell to the core. Therefore, $\tau_\text{QD}$ is given by

\noindent
\begin{equation}\label{eq:tauNR}
\frac{1}{\tau_\text{QD}} = \frac{1}{\tau_\text{r}} + \frac{1}{\tau_\text{nr}} + \frac{1}{\tau_\text{t}}.
\end{equation}

The radiative lifetime for an exciton in a QD is independent of temperature. Therefore and as shown in Fig.~\ref{fig:Figure5}(b), we can calculate the temperature dependence of  $T_\text{nr} = \left( \tau_\text{nr}^{-1} + \tau_\text{t}^{-1} \right) ^{-1}$ from Eq.~(\ref{eq:tauNR}), which is displayed in Fig.~\ref{fig:Figure5}(b). At 10~K and for an energy fluence per pulse of 0.7~\textmu J/cm$^2$, $ T_\text{nr} = 564$~ps. Since the nonradiative decay at point defects is the nonradiative channel with the smallest activation energy [Fig.~\ref{fig:Figure3}(c)], we deduce that $\tau_\text{nr} \approx 564$~ps. Since this value is much smaller than $\tau_\text{r}$, the decay of excitons in shell QDs under high excitation is dominated by nonradiative recombination already at 10~K. Between 10 and 40~K, $T_\text{nr}$ remains almost constant, indicating that $\tau_\text{nr}$ is constant as well and that $\tau_{t} \gg \tau_\text{nr}$. However, for temperatures larger than 50~K, $\tau_\text{QD}$ decreases strongly. Since the transfer of carriers to the core, with activation energy $E_{A2}$, is the dominant nonradiative mechanism for temperatures larger than 50~K [Fig.~\ref{fig:Figure3}(c)], $\tau_\text{QD} \approx \tau_\text{t}$ in this temperature range, and we find that $\tau_\text{t} < 40$~ps for temperatures higher than 100~K.

\section{Summary and conclusions}

We have described the dynamics of excitons in the shell of GaAs/(Al,Ga)As core-shell nanowires. At low temperatures, excitons are trapped in the shells by spontaneously formed quantum dots that result from inhomogeneities in the Al content. These shell QD excitons give rise to intense luminescence lines in the 1.7--1.8~eV range and exhibit a radiative lifetime of 1.7~ns. While the transfer of carriers from the shell to the core is inefficient at 10~K, the characteristic transfer time of carriers from a shell QD to the GaAs core is faster than 40~ps for temperatures larger than 100~K, leading to the suppression of the shell QD PL intensity. The carrier cooling rate in our nanowires appears to be much slower than the one in bulk GaAs, resulting in a pronounced dependence of the shell QD exciton lifetime on the density of photogenerated carriers. The origin of such the slow cooling rate of carriers in GaAs nanowires, also reported recently in Ref.~\onlinecite{Tedeschi2016} for GaAs and InP nanowires, remains to be clarified.

The very fast transfer of carriers from the shell to the core at elevated temperatures implies that the spontaneous formation of quantum dots in the (Al,Ga)As shells should not affect the room temperature operation of light-emitting diodes based on coaxial (In,Ga)As/(Al,Ga)As nanowires.\cite{Colombo2009,Dimakis2014} In contrast, at low temperatures, the shell QDs localize charge carriers very efficiently. This localization is expected to be detrimental for the formation of high-mobility electron channels along the nanowire\cite{Funk2013} as well as for fundamental studies of quantum interference effects in nanowires.\cite{Royo2015} Finally, if one desires to extend the application of shell QDs toward higher temperatures, the carrier transfer must be inhibited or at least slowed down as much as possible. A simple way to achieve this goal might be the insertion of an AlAs blocking layer in the form of an inner shell between the GaAs core and the (Al,Ga)As outer shell. 

\acknowledgments

We thank Manfred Ramsteiner for carefully reading our manuscript. P.\,C. acknowledges funding from the Fonds National Suisse de la Recherche Scientifique through project 161032. Additional funding from the Deutsche Forschungsgemeinschaft (grant Ge2224/2-2) is also acknowledged.

\bibliography{bibliography}

\begin{thebibliography}{53}%
\makeatletter
\providecommand \@ifxundefined [1]{%
 \@ifx{#1\undefined}
}%
\providecommand \@ifnum [1]{%
 \ifnum #1\expandafter \@firstoftwo
 \else \expandafter \@secondoftwo
 \fi
}%
\providecommand \@ifx [1]{%
 \ifx #1\expandafter \@firstoftwo
 \else \expandafter \@secondoftwo
 \fi
}%
\providecommand \natexlab [1]{#1}%
\providecommand \enquote  [1]{``#1''}%
\providecommand \bibnamefont  [1]{#1}%
\providecommand \bibfnamefont [1]{#1}%
\providecommand \citenamefont [1]{#1}%
\providecommand \href@noop [0]{\@secondoftwo}%
\providecommand \href [0]{\begingroup \@sanitize@url \@href}%
\providecommand \@href[1]{\@@startlink{#1}\@@href}%
\providecommand \@@href[1]{\endgroup#1\@@endlink}%
\providecommand \@sanitize@url [0]{\catcode `\\12\catcode `\$12\catcode
  `\&12\catcode `\#12\catcode `\^12\catcode `\_12\catcode `\%12\relax}%
\providecommand \@@startlink[1]{}%
\providecommand \@@endlink[0]{}%
\providecommand \url  [0]{\begingroup\@sanitize@url \@url }%
\providecommand \@url [1]{\endgroup\@href {#1}{\urlprefix }}%
\providecommand \urlprefix  [0]{URL }%
\providecommand \Eprint [0]{\href }%
\providecommand \doibase [0]{http://dx.doi.org/}%
\providecommand \selectlanguage [0]{\@gobble}%
\providecommand \bibinfo  [0]{\@secondoftwo}%
\providecommand \bibfield  [0]{\@secondoftwo}%
\providecommand \translation [1]{[#1]}%
\providecommand \BibitemOpen [0]{}%
\providecommand \bibitemStop [0]{}%
\providecommand \bibitemNoStop [0]{.\EOS\space}%
\providecommand \EOS [0]{\spacefactor3000\relax}%
\providecommand \BibitemShut  [1]{\csname bibitem#1\endcsname}%
\let\auto@bib@innerbib\@empty
\bibitem [{\citenamefont {Ohnesorge}\ \emph {et~al.}(1996)\citenamefont
  {Ohnesorge}, \citenamefont {Albrecht}, \citenamefont {Oshinowo},
  \citenamefont {Forchel},\ and\ \citenamefont {Arakawa}}]{Ohnesorge1996}%
  \BibitemOpen
  \bibfield  {author} {\bibinfo {author} {\bibfnamefont {B.}~\bibnamefont
  {Ohnesorge}}, \bibinfo {author} {\bibfnamefont {M.}~\bibnamefont {Albrecht}},
  \bibinfo {author} {\bibfnamefont {J.}~\bibnamefont {Oshinowo}}, \bibinfo
  {author} {\bibfnamefont {A.}~\bibnamefont {Forchel}}, \ and\ \bibinfo
  {author} {\bibfnamefont {Y.}~\bibnamefont {Arakawa}},\ }\href {\doibase
  10.1103/PhysRevB.54.11532} {\bibfield  {journal} {\bibinfo  {journal} {Phys.
  Rev. B}\ }\textbf {\bibinfo {volume} {54}},\ \bibinfo {pages} {11532}
  (\bibinfo {year} {1996})}\BibitemShut {NoStop}%
\bibitem [{\citenamefont {Merano}\ \emph {et~al.}(2005)\citenamefont {Merano},
  \citenamefont {Sonderegger}, \citenamefont {Crottini}, \citenamefont
  {Collin}, \citenamefont {Pelucchi}, \citenamefont {Malko}, \citenamefont
  {Baier}, \citenamefont {Kapon}, \citenamefont {Deveaud},\ and\ \citenamefont
  {Gani{\`{e}}re}}]{Merano2005}%
  \BibitemOpen
  \bibfield  {author} {\bibinfo {author} {\bibfnamefont {M.}~\bibnamefont
  {Merano}}, \bibinfo {author} {\bibfnamefont {S.}~\bibnamefont {Sonderegger}},
  \bibinfo {author} {\bibfnamefont {A.}~\bibnamefont {Crottini}}, \bibinfo
  {author} {\bibfnamefont {S.}~\bibnamefont {Collin}}, \bibinfo {author}
  {\bibfnamefont {E.}~\bibnamefont {Pelucchi}}, \bibinfo {author}
  {\bibfnamefont {A.}~\bibnamefont {Malko}}, \bibinfo {author} {\bibfnamefont
  {M.~H.}\ \bibnamefont {Baier}}, \bibinfo {author} {\bibfnamefont
  {E.}~\bibnamefont {Kapon}}, \bibinfo {author} {\bibfnamefont
  {B.}~\bibnamefont {Deveaud}}, \ and\ \bibinfo {author} {\bibfnamefont
  {J.-D.}\ \bibnamefont {Gani{\`{e}}re}},\ }\href {\doibase
  10.1038/nature04298} {\bibfield  {journal} {\bibinfo  {journal} {Nature}\
  }\textbf {\bibinfo {volume} {438}},\ \bibinfo {pages} {479} (\bibinfo {year}
  {2005})}\BibitemShut {NoStop}%
\bibitem [{\citenamefont {Ledentsov}\ \emph {et~al.}(1996)\citenamefont
  {Ledentsov}, \citenamefont {Shchukin}, \citenamefont {Grundmann},
  \citenamefont {Kirstaedter}, \citenamefont {B\"ohrer}, \citenamefont
  {Schmidt}, \citenamefont {Bimberg}, \citenamefont {Ustinov}, \citenamefont
  {Egorov}, \citenamefont {Zhukov}, \citenamefont {Kop'ev}, \citenamefont
  {Zaitsev}, \citenamefont {Gordeev}, \citenamefont {Alferov}, \citenamefont
  {Borovkov}, \citenamefont {Kosogov}, \citenamefont {Ruvimov}, \citenamefont
  {Werner}, \citenamefont {G\"osele},\ and\ \citenamefont
  {Heydenreich}}]{Ledentsov1996}%
  \BibitemOpen
  \bibfield  {author} {\bibinfo {author} {\bibfnamefont {N.~N.}\ \bibnamefont
  {Ledentsov}}, \bibinfo {author} {\bibfnamefont {V.~A.}\ \bibnamefont
  {Shchukin}}, \bibinfo {author} {\bibfnamefont {M.}~\bibnamefont {Grundmann}},
  \bibinfo {author} {\bibfnamefont {N.}~\bibnamefont {Kirstaedter}}, \bibinfo
  {author} {\bibfnamefont {J.}~\bibnamefont {B\"ohrer}}, \bibinfo {author}
  {\bibfnamefont {O.}~\bibnamefont {Schmidt}}, \bibinfo {author} {\bibfnamefont
  {D.}~\bibnamefont {Bimberg}}, \bibinfo {author} {\bibfnamefont {V.~M.}\
  \bibnamefont {Ustinov}}, \bibinfo {author} {\bibfnamefont {A.~Y.}\
  \bibnamefont {Egorov}}, \bibinfo {author} {\bibfnamefont {A.~E.}\
  \bibnamefont {Zhukov}}, \bibinfo {author} {\bibfnamefont {P.~S.}\
  \bibnamefont {Kop'ev}}, \bibinfo {author} {\bibfnamefont {S.~V.}\
  \bibnamefont {Zaitsev}}, \bibinfo {author} {\bibfnamefont {N.~Y.}\
  \bibnamefont {Gordeev}}, \bibinfo {author} {\bibfnamefont {Z.~I.}\
  \bibnamefont {Alferov}}, \bibinfo {author} {\bibfnamefont {A.~I.}\
  \bibnamefont {Borovkov}}, \bibinfo {author} {\bibfnamefont {A.~O.}\
  \bibnamefont {Kosogov}}, \bibinfo {author} {\bibfnamefont {S.~S.}\
  \bibnamefont {Ruvimov}}, \bibinfo {author} {\bibfnamefont {P.}~\bibnamefont
  {Werner}}, \bibinfo {author} {\bibfnamefont {U.}~\bibnamefont {G\"osele}}, \
  and\ \bibinfo {author} {\bibfnamefont {J.}~\bibnamefont {Heydenreich}},\
  }\href {\doibase 10.1103/PhysRevB.54.8743} {\bibfield  {journal} {\bibinfo
  {journal} {Phys. Rev. B}\ }\textbf {\bibinfo {volume} {54}},\ \bibinfo
  {pages} {8743} (\bibinfo {year} {1996})}\BibitemShut {NoStop}%
\bibitem [{\citenamefont {Michler}\ \emph {et~al.}(2000)\citenamefont
  {Michler}, \citenamefont {Kiraz}, \citenamefont {Becher}, \citenamefont
  {Schoenfeld}, \citenamefont {Petroff}, \citenamefont {Zhang}, \citenamefont
  {Hu},\ and\ \citenamefont {Imamoglu}}]{Michler2000}%
  \BibitemOpen
  \bibfield  {author} {\bibinfo {author} {\bibfnamefont {P.}~\bibnamefont
  {Michler}}, \bibinfo {author} {\bibfnamefont {A.}~\bibnamefont {Kiraz}},
  \bibinfo {author} {\bibfnamefont {C.}~\bibnamefont {Becher}}, \bibinfo
  {author} {\bibfnamefont {W.~V.}\ \bibnamefont {Schoenfeld}}, \bibinfo
  {author} {\bibfnamefont {P.~M.}\ \bibnamefont {Petroff}}, \bibinfo {author}
  {\bibfnamefont {L.}~\bibnamefont {Zhang}}, \bibinfo {author} {\bibfnamefont
  {E.}~\bibnamefont {Hu}}, \ and\ \bibinfo {author} {\bibfnamefont
  {A.}~\bibnamefont {Imamoglu}},\ }\href {\doibase
  10.1126/science.290.5500.2282} {\bibfield  {journal} {\bibinfo  {journal}
  {Science}\ }\textbf {\bibinfo {volume} {290}},\ \bibinfo {pages} {2282}
  (\bibinfo {year} {2000})}\BibitemShut {NoStop}%
\bibitem [{\citenamefont {Stevenson}\ \emph {et~al.}(2006)\citenamefont
  {Stevenson}, \citenamefont {Young}, \citenamefont {Atkinson}, \citenamefont
  {Cooper}, \citenamefont {Ritchie},\ and\ \citenamefont
  {Shields}}]{Stevenson2006}%
  \BibitemOpen
  \bibfield  {author} {\bibinfo {author} {\bibfnamefont {R.~M.}\ \bibnamefont
  {Stevenson}}, \bibinfo {author} {\bibfnamefont {R.~J.}\ \bibnamefont
  {Young}}, \bibinfo {author} {\bibfnamefont {P.}~\bibnamefont {Atkinson}},
  \bibinfo {author} {\bibfnamefont {K.}~\bibnamefont {Cooper}}, \bibinfo
  {author} {\bibfnamefont {D.~A.}\ \bibnamefont {Ritchie}}, \ and\ \bibinfo
  {author} {\bibfnamefont {A.~J.}\ \bibnamefont {Shields}},\ }\href {\doibase
  10.1038/nature04446} {\bibfield  {journal} {\bibinfo  {journal} {Nature}\
  }\textbf {\bibinfo {volume} {439}},\ \bibinfo {pages} {179} (\bibinfo {year}
  {2006})}\BibitemShut {NoStop}%
\bibitem [{\citenamefont {Brusaferri}\ \emph {et~al.}(1996)\citenamefont
  {Brusaferri}, \citenamefont {Sanguinetti}, \citenamefont {Grilli},
  \citenamefont {Guzzi}, \citenamefont {Bignazzi}, \citenamefont {Bogani},
  \citenamefont {Carraresi}, \citenamefont {Colocci}, \citenamefont {Bosacchi},
  \citenamefont {Frigeri},\ and\ \citenamefont {Franchi}}]{Brusaferri1996}%
  \BibitemOpen
  \bibfield  {author} {\bibinfo {author} {\bibfnamefont {L.}~\bibnamefont
  {Brusaferri}}, \bibinfo {author} {\bibfnamefont {S.}~\bibnamefont
  {Sanguinetti}}, \bibinfo {author} {\bibfnamefont {E.}~\bibnamefont {Grilli}},
  \bibinfo {author} {\bibfnamefont {M.}~\bibnamefont {Guzzi}}, \bibinfo
  {author} {\bibfnamefont {A.}~\bibnamefont {Bignazzi}}, \bibinfo {author}
  {\bibfnamefont {F.}~\bibnamefont {Bogani}}, \bibinfo {author} {\bibfnamefont
  {L.}~\bibnamefont {Carraresi}}, \bibinfo {author} {\bibfnamefont
  {M.}~\bibnamefont {Colocci}}, \bibinfo {author} {\bibfnamefont
  {A.}~\bibnamefont {Bosacchi}}, \bibinfo {author} {\bibfnamefont
  {P.}~\bibnamefont {Frigeri}}, \ and\ \bibinfo {author} {\bibfnamefont
  {S.}~\bibnamefont {Franchi}},\ }\href {\doibase
  http://dx.doi.org/10.1063/1.117304} {\bibfield  {journal} {\bibinfo
  {journal} {Appl. Phys. Lett.}\ }\textbf {\bibinfo {volume} {69}},\ \bibinfo
  {pages} {3354} (\bibinfo {year} {1996})}\BibitemShut {NoStop}%
\bibitem [{\citenamefont {Findeis}\ \emph {et~al.}(2001)\citenamefont
  {Findeis}, \citenamefont {Baier}, \citenamefont {Beham}, \citenamefont
  {Zrenner},\ and\ \citenamefont {Abstreiter}}]{Findeis2001}%
  \BibitemOpen
  \bibfield  {author} {\bibinfo {author} {\bibfnamefont {F.}~\bibnamefont
  {Findeis}}, \bibinfo {author} {\bibfnamefont {M.}~\bibnamefont {Baier}},
  \bibinfo {author} {\bibfnamefont {E.}~\bibnamefont {Beham}}, \bibinfo
  {author} {\bibfnamefont {A.}~\bibnamefont {Zrenner}}, \ and\ \bibinfo
  {author} {\bibfnamefont {G.}~\bibnamefont {Abstreiter}},\ }\href {\doibase
  http://dx.doi.org/10.1063/1.1369148} {\bibfield  {journal} {\bibinfo
  {journal} {Appl. Phys. Lett.}\ }\textbf {\bibinfo {volume} {78}},\ \bibinfo
  {pages} {2958} (\bibinfo {year} {2001})}\BibitemShut {NoStop}%
\bibitem [{\citenamefont {Smith}\ \emph {et~al.}(2005)\citenamefont {Smith},
  \citenamefont {Dalgarno}, \citenamefont {Warburton}, \citenamefont {Govorov},
  \citenamefont {Karrai}, \citenamefont {Gerardot},\ and\ \citenamefont
  {Petroff}}]{Smith2005}%
  \BibitemOpen
  \bibfield  {author} {\bibinfo {author} {\bibfnamefont {J.~M.}\ \bibnamefont
  {Smith}}, \bibinfo {author} {\bibfnamefont {P.~A.}\ \bibnamefont {Dalgarno}},
  \bibinfo {author} {\bibfnamefont {R.~J.}\ \bibnamefont {Warburton}}, \bibinfo
  {author} {\bibfnamefont {A.~O.}\ \bibnamefont {Govorov}}, \bibinfo {author}
  {\bibfnamefont {K.}~\bibnamefont {Karrai}}, \bibinfo {author} {\bibfnamefont
  {B.~D.}\ \bibnamefont {Gerardot}}, \ and\ \bibinfo {author} {\bibfnamefont
  {P.~M.}\ \bibnamefont {Petroff}},\ }\href {\doibase
  10.1103/PhysRevLett.94.197402} {\bibfield  {journal} {\bibinfo  {journal}
  {Phys. Rev. Lett.}\ }\textbf {\bibinfo {volume} {94}},\ \bibinfo {pages}
  {197402} (\bibinfo {year} {2005})}\BibitemShut {NoStop}%
\bibitem [{\citenamefont {Van~Hattem}\ \emph {et~al.}(2013)\citenamefont
  {Van~Hattem}, \citenamefont {Corfdir}, \citenamefont {Brereton},
  \citenamefont {Pearce}, \citenamefont {Graham}, \citenamefont {Stanley},
  \citenamefont {Hugues}, \citenamefont {Hopkinson},\ and\ \citenamefont
  {Phillips}}]{VanHattem2013}%
  \BibitemOpen
  \bibfield  {author} {\bibinfo {author} {\bibfnamefont {B.}~\bibnamefont
  {Van~Hattem}}, \bibinfo {author} {\bibfnamefont {P.}~\bibnamefont {Corfdir}},
  \bibinfo {author} {\bibfnamefont {P.}~\bibnamefont {Brereton}}, \bibinfo
  {author} {\bibfnamefont {P.}~\bibnamefont {Pearce}}, \bibinfo {author}
  {\bibfnamefont {A.~M.}\ \bibnamefont {Graham}}, \bibinfo {author}
  {\bibfnamefont {M.~J.}\ \bibnamefont {Stanley}}, \bibinfo {author}
  {\bibfnamefont {M.}~\bibnamefont {Hugues}}, \bibinfo {author} {\bibfnamefont
  {M.}~\bibnamefont {Hopkinson}}, \ and\ \bibinfo {author} {\bibfnamefont
  {R.~T.}\ \bibnamefont {Phillips}},\ }\href {\doibase
  10.1103/PhysRevB.87.205308} {\bibfield  {journal} {\bibinfo  {journal} {Phys.
  Rev. B}\ }\textbf {\bibinfo {volume} {87}},\ \bibinfo {pages} {205308}
  (\bibinfo {year} {2013})}\BibitemShut {NoStop}%
\bibitem [{\citenamefont {Warburton}\ \emph {et~al.}(2000)\citenamefont
  {Warburton}, \citenamefont {Sch{\"{a}}flein}, \citenamefont {Haft},
  \citenamefont {Bickel}, \citenamefont {Lorke}, \citenamefont {Karrai},
  \citenamefont {Garcia}, \citenamefont {Schoenfeld},\ and\ \citenamefont
  {Petroff}}]{Warburton2000}%
  \BibitemOpen
  \bibfield  {author} {\bibinfo {author} {\bibfnamefont {R.~J.}\ \bibnamefont
  {Warburton}}, \bibinfo {author} {\bibfnamefont {C.}~\bibnamefont
  {Sch{\"{a}}flein}}, \bibinfo {author} {\bibfnamefont {D.}~\bibnamefont
  {Haft}}, \bibinfo {author} {\bibfnamefont {F.}~\bibnamefont {Bickel}},
  \bibinfo {author} {\bibfnamefont {A.}~\bibnamefont {Lorke}}, \bibinfo
  {author} {\bibfnamefont {K.}~\bibnamefont {Karrai}}, \bibinfo {author}
  {\bibfnamefont {J.~M.}\ \bibnamefont {Garcia}}, \bibinfo {author}
  {\bibfnamefont {W.}~\bibnamefont {Schoenfeld}}, \ and\ \bibinfo {author}
  {\bibfnamefont {P.~M.}\ \bibnamefont {Petroff}},\ }\href {\doibase
  10.1038/35016030} {\bibfield  {journal} {\bibinfo  {journal} {Nature}\
  }\textbf {\bibinfo {volume} {405}},\ \bibinfo {pages} {926} (\bibinfo {year}
  {2000})}\BibitemShut {NoStop}%
\bibitem [{\citenamefont {Heiss}\ \emph {et~al.}(2013)\citenamefont {Heiss},
  \citenamefont {Fontana}, \citenamefont {Gustafsson}, \citenamefont
  {W{\"{u}}st}, \citenamefont {Magen}, \citenamefont {O’Regan}, \citenamefont
  {Luo}, \citenamefont {Ketterer}, \citenamefont {Conesa-Boj}, \citenamefont
  {Kuhlmann}, \citenamefont {Houel}, \citenamefont {Russo-Averchi},
  \citenamefont {Morante}, \citenamefont {Cantoni}, \citenamefont {Marzari},
  \citenamefont {Arbiol}, \citenamefont {Zunger}, \citenamefont {Warburton},\
  and\ \citenamefont {{Fontcuberta i Morral}}}]{Heiss2013}%
  \BibitemOpen
  \bibfield  {author} {\bibinfo {author} {\bibfnamefont {M.}~\bibnamefont
  {Heiss}}, \bibinfo {author} {\bibfnamefont {Y.}~\bibnamefont {Fontana}},
  \bibinfo {author} {\bibfnamefont {A.}~\bibnamefont {Gustafsson}}, \bibinfo
  {author} {\bibfnamefont {G.}~\bibnamefont {W{\"{u}}st}}, \bibinfo {author}
  {\bibfnamefont {C.}~\bibnamefont {Magen}}, \bibinfo {author} {\bibfnamefont
  {D.~D.}\ \bibnamefont {O’Regan}}, \bibinfo {author} {\bibfnamefont {J.~W.}\
  \bibnamefont {Luo}}, \bibinfo {author} {\bibfnamefont {B.}~\bibnamefont
  {Ketterer}}, \bibinfo {author} {\bibfnamefont {S.}~\bibnamefont
  {Conesa-Boj}}, \bibinfo {author} {\bibfnamefont {A.~V.}\ \bibnamefont
  {Kuhlmann}}, \bibinfo {author} {\bibfnamefont {J.}~\bibnamefont {Houel}},
  \bibinfo {author} {\bibfnamefont {E.}~\bibnamefont {Russo-Averchi}}, \bibinfo
  {author} {\bibfnamefont {J.~R.}\ \bibnamefont {Morante}}, \bibinfo {author}
  {\bibfnamefont {M.}~\bibnamefont {Cantoni}}, \bibinfo {author} {\bibfnamefont
  {N.}~\bibnamefont {Marzari}}, \bibinfo {author} {\bibfnamefont
  {J.}~\bibnamefont {Arbiol}}, \bibinfo {author} {\bibfnamefont
  {A.}~\bibnamefont {Zunger}}, \bibinfo {author} {\bibfnamefont {R.~J.}\
  \bibnamefont {Warburton}}, \ and\ \bibinfo {author} {\bibfnamefont
  {A.}~\bibnamefont {{Fontcuberta i Morral}}},\ }\href {\doibase
  10.1038/nmat3557} {\bibfield  {journal} {\bibinfo  {journal} {Nat. Mater.}\
  }\textbf {\bibinfo {volume} {12}},\ \bibinfo {pages} {439} (\bibinfo {year}
  {2013})}\BibitemShut {NoStop}%
\bibitem [{\citenamefont {Rudolph}\ \emph
  {et~al.}(2013{\natexlab{a}})\citenamefont {Rudolph}, \citenamefont {Funk},
  \citenamefont {D{\"{o}}blinger}, \citenamefont {Mork{\"{o}}tter},
  \citenamefont {Hertenberger}, \citenamefont {Schweickert}, \citenamefont
  {Becker}, \citenamefont {Matich}, \citenamefont {Bichler}, \citenamefont
  {Spirkoska}, \citenamefont {Zardo}, \citenamefont {Finley}, \citenamefont
  {Abstreiter},\ and\ \citenamefont {Koblm{\"{u}}ller}}]{Rudolph2013a}%
  \BibitemOpen
  \bibfield  {author} {\bibinfo {author} {\bibfnamefont {D.}~\bibnamefont
  {Rudolph}}, \bibinfo {author} {\bibfnamefont {S.}~\bibnamefont {Funk}},
  \bibinfo {author} {\bibfnamefont {M.}~\bibnamefont {D{\"{o}}blinger}},
  \bibinfo {author} {\bibfnamefont {S.}~\bibnamefont {Mork{\"{o}}tter}},
  \bibinfo {author} {\bibfnamefont {S.}~\bibnamefont {Hertenberger}}, \bibinfo
  {author} {\bibfnamefont {L.}~\bibnamefont {Schweickert}}, \bibinfo {author}
  {\bibfnamefont {J.}~\bibnamefont {Becker}}, \bibinfo {author} {\bibfnamefont
  {S.}~\bibnamefont {Matich}}, \bibinfo {author} {\bibfnamefont
  {M.}~\bibnamefont {Bichler}}, \bibinfo {author} {\bibfnamefont
  {D.}~\bibnamefont {Spirkoska}}, \bibinfo {author} {\bibfnamefont
  {I.}~\bibnamefont {Zardo}}, \bibinfo {author} {\bibfnamefont {J.~J.}\
  \bibnamefont {Finley}}, \bibinfo {author} {\bibfnamefont {G.}~\bibnamefont
  {Abstreiter}}, \ and\ \bibinfo {author} {\bibfnamefont {G.}~\bibnamefont
  {Koblm{\"{u}}ller}},\ }\href {\doibase 10.1021/nl3046816} {\bibfield
  {journal} {\bibinfo  {journal} {Nano Lett.}\ }\textbf {\bibinfo {volume}
  {13}},\ \bibinfo {pages} {1522} (\bibinfo {year}
  {2013}{\natexlab{a}})}\BibitemShut {NoStop}%
\bibitem [{\citenamefont {Fontana}\ \emph {et~al.}(2014)\citenamefont
  {Fontana}, \citenamefont {Corfdir}, \citenamefont {{Van Hattem}},
  \citenamefont {Russo-Averchi}, \citenamefont {Heiss}, \citenamefont
  {Sonderegger}, \citenamefont {Magen}, \citenamefont {Arbiol}, \citenamefont
  {Phillips},\ and\ \citenamefont {{Fontcuberta i Morral}}}]{Fontana2014}%
  \BibitemOpen
  \bibfield  {author} {\bibinfo {author} {\bibfnamefont {Y.}~\bibnamefont
  {Fontana}}, \bibinfo {author} {\bibfnamefont {P.}~\bibnamefont {Corfdir}},
  \bibinfo {author} {\bibfnamefont {B.}~\bibnamefont {{Van Hattem}}}, \bibinfo
  {author} {\bibfnamefont {E.}~\bibnamefont {Russo-Averchi}}, \bibinfo {author}
  {\bibfnamefont {M.}~\bibnamefont {Heiss}}, \bibinfo {author} {\bibfnamefont
  {S.}~\bibnamefont {Sonderegger}}, \bibinfo {author} {\bibfnamefont
  {C.}~\bibnamefont {Magen}}, \bibinfo {author} {\bibfnamefont
  {J.}~\bibnamefont {Arbiol}}, \bibinfo {author} {\bibfnamefont {R.~T.}\
  \bibnamefont {Phillips}}, \ and\ \bibinfo {author} {\bibfnamefont
  {A.}~\bibnamefont {{Fontcuberta i Morral}}},\ }\href {\doibase
  10.1103/PhysRevB.90.075307} {\bibfield  {journal} {\bibinfo  {journal} {Phys.
  Rev. B}\ }\textbf {\bibinfo {volume} {90}},\ \bibinfo {pages} {075307}
  (\bibinfo {year} {2014})}\BibitemShut {NoStop}%
\bibitem [{\citenamefont {Jeon}\ \emph {et~al.}(2015)\citenamefont {Jeon},
  \citenamefont {Loitsch}, \citenamefont {Morkoetter}, \citenamefont
  {Abstreiter}, \citenamefont {Finley}, \citenamefont {Krenner}, \citenamefont
  {Koblmueller},\ and\ \citenamefont {Lauhon}}]{Jeon2015}%
  \BibitemOpen
  \bibfield  {author} {\bibinfo {author} {\bibfnamefont {N.}~\bibnamefont
  {Jeon}}, \bibinfo {author} {\bibfnamefont {B.}~\bibnamefont {Loitsch}},
  \bibinfo {author} {\bibfnamefont {S.}~\bibnamefont {Morkoetter}}, \bibinfo
  {author} {\bibfnamefont {G.}~\bibnamefont {Abstreiter}}, \bibinfo {author}
  {\bibfnamefont {J.}~\bibnamefont {Finley}}, \bibinfo {author} {\bibfnamefont
  {H.~J.}\ \bibnamefont {Krenner}}, \bibinfo {author} {\bibfnamefont
  {G.}~\bibnamefont {Koblmueller}}, \ and\ \bibinfo {author} {\bibfnamefont
  {L.~J.}\ \bibnamefont {Lauhon}},\ }\href {\doibase 10.1021/acsnano.5b04070}
  {\bibfield  {journal} {\bibinfo  {journal} {ACS Nano}\ }\textbf {\bibinfo
  {volume} {9}},\ \bibinfo {pages} {8335} (\bibinfo {year} {2015})}\BibitemShut
  {NoStop}%
\bibitem [{\citenamefont {Mancini}\ \emph {et~al.}(2014)\citenamefont
  {Mancini}, \citenamefont {Fontana}, \citenamefont {Conesa-Boj}, \citenamefont
  {Blum}, \citenamefont {Vurpillot}, \citenamefont {Francaviglia},
  \citenamefont {Russo-Averchi}, \citenamefont {Heiss}, \citenamefont {Arbiol},
  \citenamefont {{Fontcuberta i Morral}},\ and\ \citenamefont
  {Rigutti}}]{Mancini2014}%
  \BibitemOpen
  \bibfield  {author} {\bibinfo {author} {\bibfnamefont {L.}~\bibnamefont
  {Mancini}}, \bibinfo {author} {\bibfnamefont {Y.}~\bibnamefont {Fontana}},
  \bibinfo {author} {\bibfnamefont {S.}~\bibnamefont {Conesa-Boj}}, \bibinfo
  {author} {\bibfnamefont {I.}~\bibnamefont {Blum}}, \bibinfo {author}
  {\bibfnamefont {F.}~\bibnamefont {Vurpillot}}, \bibinfo {author}
  {\bibfnamefont {L.}~\bibnamefont {Francaviglia}}, \bibinfo {author}
  {\bibfnamefont {E.}~\bibnamefont {Russo-Averchi}}, \bibinfo {author}
  {\bibfnamefont {M.}~\bibnamefont {Heiss}}, \bibinfo {author} {\bibfnamefont
  {J.}~\bibnamefont {Arbiol}}, \bibinfo {author} {\bibfnamefont
  {A.}~\bibnamefont {{Fontcuberta i Morral}}}, \ and\ \bibinfo {author}
  {\bibfnamefont {L.}~\bibnamefont {Rigutti}},\ }\href {\doibase
  http://dx.doi.org/10.1063/1.4904952} {\bibfield  {journal} {\bibinfo
  {journal} {Appl. Phys. Lett.}\ }\textbf {\bibinfo {volume} {105}},\ \bibinfo
  {pages} {243106} (\bibinfo {year} {2014})}\BibitemShut {NoStop}%
\bibitem [{\citenamefont {Montinaro}\ \emph {et~al.}(2014)\citenamefont
  {Montinaro}, \citenamefont {W{\"{u}}st}, \citenamefont {Munsch},
  \citenamefont {Fontana}, \citenamefont {Russo-Averchi}, \citenamefont
  {Heiss}, \citenamefont {{Fontcuberta i Morral}}, \citenamefont {Warburton},\
  and\ \citenamefont {Poggio}}]{Montinaro2014}%
  \BibitemOpen
  \bibfield  {author} {\bibinfo {author} {\bibfnamefont {M.}~\bibnamefont
  {Montinaro}}, \bibinfo {author} {\bibfnamefont {G.}~\bibnamefont
  {W{\"{u}}st}}, \bibinfo {author} {\bibfnamefont {M.}~\bibnamefont {Munsch}},
  \bibinfo {author} {\bibfnamefont {Y.}~\bibnamefont {Fontana}}, \bibinfo
  {author} {\bibfnamefont {E.}~\bibnamefont {Russo-Averchi}}, \bibinfo {author}
  {\bibfnamefont {M.}~\bibnamefont {Heiss}}, \bibinfo {author} {\bibfnamefont
  {A.}~\bibnamefont {{Fontcuberta i Morral}}}, \bibinfo {author} {\bibfnamefont
  {R.~J.}\ \bibnamefont {Warburton}}, \ and\ \bibinfo {author} {\bibfnamefont
  {M.}~\bibnamefont {Poggio}},\ }\href {\doibase 10.1021/nl501413t} {\bibfield
  {journal} {\bibinfo  {journal} {Nano Lett.}\ }\textbf {\bibinfo {volume}
  {14}},\ \bibinfo {pages} {4454} (\bibinfo {year} {2014})}\BibitemShut
  {NoStop}%
\bibitem [{\citenamefont {Graham}\ \emph {et~al.}(2013)\citenamefont {Graham},
  \citenamefont {Corfdir}, \citenamefont {Heiss}, \citenamefont {Conesa-Boj},
  \citenamefont {Uccelli}, \citenamefont {{Fontcuberta i Morral}},\ and\
  \citenamefont {Phillips}}]{Graham2013}%
  \BibitemOpen
  \bibfield  {author} {\bibinfo {author} {\bibfnamefont {A.~M.}\ \bibnamefont
  {Graham}}, \bibinfo {author} {\bibfnamefont {P.}~\bibnamefont {Corfdir}},
  \bibinfo {author} {\bibfnamefont {M.}~\bibnamefont {Heiss}}, \bibinfo
  {author} {\bibfnamefont {S.}~\bibnamefont {Conesa-Boj}}, \bibinfo {author}
  {\bibfnamefont {E.}~\bibnamefont {Uccelli}}, \bibinfo {author} {\bibfnamefont
  {A.}~\bibnamefont {{Fontcuberta i Morral}}}, \ and\ \bibinfo {author}
  {\bibfnamefont {R.~T.}\ \bibnamefont {Phillips}},\ }\href {\doibase
  10.1103/PhysRevB.87.125304} {\bibfield  {journal} {\bibinfo  {journal} {Phys.
  Rev. B}\ }\textbf {\bibinfo {volume} {87}},\ \bibinfo {pages} {125304}
  (\bibinfo {year} {2013})}\BibitemShut {NoStop}%
\bibitem [{\citenamefont {Martelli}\ \emph {et~al.}(2015)\citenamefont
  {Martelli}, \citenamefont {Priante},\ and\ \citenamefont
  {Rubini}}]{Martelli2015}%
  \BibitemOpen
  \bibfield  {author} {\bibinfo {author} {\bibfnamefont {F.}~\bibnamefont
  {Martelli}}, \bibinfo {author} {\bibfnamefont {G.}~\bibnamefont {Priante}}, \
  and\ \bibinfo {author} {\bibfnamefont {S.}~\bibnamefont {Rubini}},\ }\href
  {\doibase 10.1088/0268-1242/30/5/055020} {\bibfield  {journal} {\bibinfo
  {journal} {Semicond. Sci. Technol.}\ }\textbf {\bibinfo {volume} {30}},\
  \bibinfo {pages} {055020} (\bibinfo {year} {2015})}\BibitemShut {NoStop}%
\bibitem [{\citenamefont {Ramsteiner}\ \emph {et~al.}(1997)\citenamefont
  {Ramsteiner}, \citenamefont {Hey}, \citenamefont {Klann}, \citenamefont
  {Jahn}, \citenamefont {Gorbunova},\ and\ \citenamefont
  {Ploog}}]{Ramsteiner1997}%
  \BibitemOpen
  \bibfield  {author} {\bibinfo {author} {\bibfnamefont {M.}~\bibnamefont
  {Ramsteiner}}, \bibinfo {author} {\bibfnamefont {R.}~\bibnamefont {Hey}},
  \bibinfo {author} {\bibfnamefont {R.}~\bibnamefont {Klann}}, \bibinfo
  {author} {\bibfnamefont {U.}~\bibnamefont {Jahn}}, \bibinfo {author}
  {\bibfnamefont {I.}~\bibnamefont {Gorbunova}}, \ and\ \bibinfo {author}
  {\bibfnamefont {K.~H.}\ \bibnamefont {Ploog}},\ }\href {\doibase
  10.1103/PhysRevB.55.5239} {\bibfield  {journal} {\bibinfo  {journal} {Phys.
  Rev. B}\ }\textbf {\bibinfo {volume} {55}},\ \bibinfo {pages} {5239}
  (\bibinfo {year} {1997})}\BibitemShut {NoStop}%
\bibitem [{\citenamefont {Wei{\ss}}\ \emph {et~al.}(2014)\citenamefont
  {Wei{\ss}}, \citenamefont {Kinzel}, \citenamefont {Sch{\"{u}}lein},
  \citenamefont {Heigl}, \citenamefont {Rudolph}, \citenamefont
  {Mork{\"{o}}tter}, \citenamefont {D{\"{o}}blinger}, \citenamefont {Bichler},
  \citenamefont {Abstreiter}, \citenamefont {Finley}, \citenamefont
  {Koblm{\"{u}}ller}, \citenamefont {Wixforth},\ and\ \citenamefont
  {Krenner}}]{Weiss2014}%
  \BibitemOpen
  \bibfield  {author} {\bibinfo {author} {\bibfnamefont {M.}~\bibnamefont
  {Wei{\ss}}}, \bibinfo {author} {\bibfnamefont {J.~B.}\ \bibnamefont
  {Kinzel}}, \bibinfo {author} {\bibfnamefont {F.~J.~R.}\ \bibnamefont
  {Sch{\"{u}}lein}}, \bibinfo {author} {\bibfnamefont {M.}~\bibnamefont
  {Heigl}}, \bibinfo {author} {\bibfnamefont {D.}~\bibnamefont {Rudolph}},
  \bibinfo {author} {\bibfnamefont {S.}~\bibnamefont {Mork{\"{o}}tter}},
  \bibinfo {author} {\bibfnamefont {M.}~\bibnamefont {D{\"{o}}blinger}},
  \bibinfo {author} {\bibfnamefont {M.}~\bibnamefont {Bichler}}, \bibinfo
  {author} {\bibfnamefont {G.}~\bibnamefont {Abstreiter}}, \bibinfo {author}
  {\bibfnamefont {J.~J.}\ \bibnamefont {Finley}}, \bibinfo {author}
  {\bibfnamefont {G.}~\bibnamefont {Koblm{\"{u}}ller}}, \bibinfo {author}
  {\bibfnamefont {A.}~\bibnamefont {Wixforth}}, \ and\ \bibinfo {author}
  {\bibfnamefont {H.~J.}\ \bibnamefont {Krenner}},\ }\href {\doibase
  10.1021/nl4040434} {\bibfield  {journal} {\bibinfo  {journal} {Nano Lett.}\
  }\textbf {\bibinfo {volume} {14}},\ \bibinfo {pages} {2256} (\bibinfo {year}
  {2014})}\BibitemShut {NoStop}%
\bibitem [{\citenamefont {Francaviglia}\ \emph {et~al.}(2015)\citenamefont
  {Francaviglia}, \citenamefont {Fontana}, \citenamefont {Conesa-Boj},
  \citenamefont {T{\"{u}}t{\"{u}}nc{\"{u}}oglu}, \citenamefont {Duch{\^{e}}ne},
  \citenamefont {Tanasescu}, \citenamefont {Matteini},\ and\ \citenamefont
  {{Fontcuberta i Morral}}}]{Francaviglia2015}%
  \BibitemOpen
  \bibfield  {author} {\bibinfo {author} {\bibfnamefont {L.}~\bibnamefont
  {Francaviglia}}, \bibinfo {author} {\bibfnamefont {Y.}~\bibnamefont
  {Fontana}}, \bibinfo {author} {\bibfnamefont {S.}~\bibnamefont {Conesa-Boj}},
  \bibinfo {author} {\bibfnamefont {G.}~\bibnamefont
  {T{\"{u}}t{\"{u}}nc{\"{u}}oglu}}, \bibinfo {author} {\bibfnamefont
  {L.}~\bibnamefont {Duch{\^{e}}ne}}, \bibinfo {author} {\bibfnamefont {M.~B.}\
  \bibnamefont {Tanasescu}}, \bibinfo {author} {\bibfnamefont {F.}~\bibnamefont
  {Matteini}}, \ and\ \bibinfo {author} {\bibfnamefont {A.}~\bibnamefont
  {{Fontcuberta i Morral}}},\ }\href {\doibase 10.1063/1.4927315} {\bibfield
  {journal} {\bibinfo  {journal} {Appl. Phys. Lett.}\ }\textbf {\bibinfo
  {volume} {107}},\ \bibinfo {pages} {033106} (\bibinfo {year}
  {2015})}\BibitemShut {NoStop}%
\bibitem [{\citenamefont {Corfdir}\ \emph {et~al.}(2014)\citenamefont
  {Corfdir}, \citenamefont {Fontana}, \citenamefont {{Van Hattem}},
  \citenamefont {Russo-Averchi}, \citenamefont {Heiss}, \citenamefont
  {{Fontcuberta i Morral}},\ and\ \citenamefont {Phillips}}]{Corfdir2014}%
  \BibitemOpen
  \bibfield  {author} {\bibinfo {author} {\bibfnamefont {P.}~\bibnamefont
  {Corfdir}}, \bibinfo {author} {\bibfnamefont {Y.}~\bibnamefont {Fontana}},
  \bibinfo {author} {\bibfnamefont {B.}~\bibnamefont {{Van Hattem}}}, \bibinfo
  {author} {\bibfnamefont {E.}~\bibnamefont {Russo-Averchi}}, \bibinfo {author}
  {\bibfnamefont {M.}~\bibnamefont {Heiss}}, \bibinfo {author} {\bibfnamefont
  {A.}~\bibnamefont {{Fontcuberta i Morral}}}, \ and\ \bibinfo {author}
  {\bibfnamefont {R.~T.}\ \bibnamefont {Phillips}},\ }\href {\doibase
  10.1063/1.4903515} {\bibfield  {journal} {\bibinfo  {journal} {Appl. Phys.
  Lett.}\ }\textbf {\bibinfo {volume} {105}},\ \bibinfo {pages} {223111}
  (\bibinfo {year} {2014})}\BibitemShut {NoStop}%
\bibitem [{\citenamefont {Holmes}\ \emph {et~al.}(2015)\citenamefont {Holmes},
  \citenamefont {Kako}, \citenamefont {Choi}, \citenamefont {Arita},\ and\
  \citenamefont {Arakawa}}]{Holmes2015}%
  \BibitemOpen
  \bibfield  {author} {\bibinfo {author} {\bibfnamefont {M.}~\bibnamefont
  {Holmes}}, \bibinfo {author} {\bibfnamefont {S.}~\bibnamefont {Kako}},
  \bibinfo {author} {\bibfnamefont {K.}~\bibnamefont {Choi}}, \bibinfo {author}
  {\bibfnamefont {M.}~\bibnamefont {Arita}}, \ and\ \bibinfo {author}
  {\bibfnamefont {Y.}~\bibnamefont {Arakawa}},\ }\href {\doibase
  10.1103/PhysRevB.92.115447} {\bibfield  {journal} {\bibinfo  {journal} {Phys.
  Rev. B}\ }\textbf {\bibinfo {volume} {92}},\ \bibinfo {pages} {115447}
  (\bibinfo {year} {2015})}\BibitemShut {NoStop}%
\bibitem [{\citenamefont {Langbein}\ and\ \citenamefont
  {Hvam}(1999)}]{Langbein1999}%
  \BibitemOpen
  \bibfield  {author} {\bibinfo {author} {\bibfnamefont {W.}~\bibnamefont
  {Langbein}}\ and\ \bibinfo {author} {\bibfnamefont {J.~M.}\ \bibnamefont
  {Hvam}},\ }\href {\doibase 10.1103/PhysRevB.59.15405} {\bibfield  {journal}
  {\bibinfo  {journal} {Phys. Rev. B}\ }\textbf {\bibinfo {volume} {59}},\
  \bibinfo {pages} {15405} (\bibinfo {year} {1999})}\BibitemShut {NoStop}%
\bibitem [{\citenamefont {Seguin}\ \emph {et~al.}(2005)\citenamefont {Seguin},
  \citenamefont {Schliwa}, \citenamefont {Rodt}, \citenamefont {P\"otschke},
  \citenamefont {Pohl},\ and\ \citenamefont {Bimberg}}]{Seguin2005}%
  \BibitemOpen
  \bibfield  {author} {\bibinfo {author} {\bibfnamefont {R.}~\bibnamefont
  {Seguin}}, \bibinfo {author} {\bibfnamefont {A.}~\bibnamefont {Schliwa}},
  \bibinfo {author} {\bibfnamefont {S.}~\bibnamefont {Rodt}}, \bibinfo {author}
  {\bibfnamefont {K.}~\bibnamefont {P\"otschke}}, \bibinfo {author}
  {\bibfnamefont {U.~W.}\ \bibnamefont {Pohl}}, \ and\ \bibinfo {author}
  {\bibfnamefont {D.}~\bibnamefont {Bimberg}},\ }\href {\doibase
  10.1103/PhysRevLett.95.257402} {\bibfield  {journal} {\bibinfo  {journal}
  {Phys. Rev. Lett.}\ }\textbf {\bibinfo {volume} {95}},\ \bibinfo {pages}
  {257402} (\bibinfo {year} {2005})}\BibitemShut {NoStop}%
\bibitem [{\citenamefont {Finley}\ \emph {et~al.}(2002)\citenamefont {Finley},
  \citenamefont {Mowbray}, \citenamefont {Skolnick}, \citenamefont {Ashmore},
  \citenamefont {Baker}, \citenamefont {Monte},\ and\ \citenamefont
  {Hopkinson}}]{Finley2002}%
  \BibitemOpen
  \bibfield  {author} {\bibinfo {author} {\bibfnamefont {J.~J.}\ \bibnamefont
  {Finley}}, \bibinfo {author} {\bibfnamefont {D.~J.}\ \bibnamefont {Mowbray}},
  \bibinfo {author} {\bibfnamefont {M.~S.}\ \bibnamefont {Skolnick}}, \bibinfo
  {author} {\bibfnamefont {A.~D.}\ \bibnamefont {Ashmore}}, \bibinfo {author}
  {\bibfnamefont {C.}~\bibnamefont {Baker}}, \bibinfo {author} {\bibfnamefont
  {A.~F.~G.}\ \bibnamefont {Monte}}, \ and\ \bibinfo {author} {\bibfnamefont
  {M.}~\bibnamefont {Hopkinson}},\ }\href {\doibase 10.1103/PhysRevB.66.153316}
  {\bibfield  {journal} {\bibinfo  {journal} {Phys. Rev. B}\ }\textbf {\bibinfo
  {volume} {66}},\ \bibinfo {pages} {153316} (\bibinfo {year}
  {2002})}\BibitemShut {NoStop}%
\bibitem [{\citenamefont {O'Donnell}\ \emph {et~al.}(1999)\citenamefont
  {O'Donnell}, \citenamefont {Martin},\ and\ \citenamefont
  {Middleton}}]{ODonnell1999}%
  \BibitemOpen
  \bibfield  {author} {\bibinfo {author} {\bibfnamefont {K.~P.}\ \bibnamefont
  {O'Donnell}}, \bibinfo {author} {\bibfnamefont {R.~W.}\ \bibnamefont
  {Martin}}, \ and\ \bibinfo {author} {\bibfnamefont {P.~G.}\ \bibnamefont
  {Middleton}},\ }\href {\doibase 10.1103/PhysRevLett.82.237} {\bibfield
  {journal} {\bibinfo  {journal} {Phys. Rev. Lett.}\ }\textbf {\bibinfo
  {volume} {82}},\ \bibinfo {pages} {237} (\bibinfo {year} {1999})}\BibitemShut
  {NoStop}%
\bibitem [{\citenamefont {Corfdir}\ \emph {et~al.}(2015)\citenamefont
  {Corfdir}, \citenamefont {Feix}, \citenamefont {Zettler}, \citenamefont
  {Fern{\'{a}}ndez-Garrido},\ and\ \citenamefont {Brandt}}]{Corfdir2015}%
  \BibitemOpen
  \bibfield  {author} {\bibinfo {author} {\bibfnamefont {P.}~\bibnamefont
  {Corfdir}}, \bibinfo {author} {\bibfnamefont {F.}~\bibnamefont {Feix}},
  \bibinfo {author} {\bibfnamefont {J.~K.}\ \bibnamefont {Zettler}}, \bibinfo
  {author} {\bibfnamefont {S.}~\bibnamefont {Fern{\'{a}}ndez-Garrido}}, \ and\
  \bibinfo {author} {\bibfnamefont {O.}~\bibnamefont {Brandt}},\ }\href
  {\doibase 10.1088/1367-2630/17/3/033040} {\bibfield  {journal} {\bibinfo
  {journal} {New J. Phys.}\ }\textbf {\bibinfo {volume} {17}},\ \bibinfo
  {pages} {033040} (\bibinfo {year} {2015})}\BibitemShut {NoStop}%
\bibitem [{\citenamefont {P{\"{a}}ssler}(1997)}]{Passler1997}%
  \BibitemOpen
  \bibfield  {author} {\bibinfo {author} {\bibfnamefont {R.}~\bibnamefont
  {P{\"{a}}ssler}},\ }\href {\doibase DOI:
  10.1002/1521-3951(199703)200:1<155::AID-PSSB155>3.0.CO;2-3} {\bibfield
  {journal} {\bibinfo  {journal} {Phys. Status Solidi B}\ }\textbf {\bibinfo
  {volume} {200}},\ \bibinfo {pages} {155} (\bibinfo {year}
  {1997})}\BibitemShut {NoStop}%
\bibitem [{\citenamefont {P{\"{a}}ssler}(1999)}]{Passler1999}%
  \BibitemOpen
  \bibfield  {author} {\bibinfo {author} {\bibfnamefont {R.}~\bibnamefont
  {P{\"{a}}ssler}},\ }\href {\doibase
  10.1002/(SICI)1521-3951(199912)216:2<975::AID-PSSB975>3.0.CO;2-N} {\bibfield
  {journal} {\bibinfo  {journal} {Phys. Status Solidi B}\ }\textbf {\bibinfo
  {volume} {216}},\ \bibinfo {pages} {975} (\bibinfo {year}
  {1999})}\BibitemShut {NoStop}%
\bibitem [{\citenamefont {Zilli}\ \emph {et~al.}(2015)\citenamefont {Zilli},
  \citenamefont {Luca}, \citenamefont {Tedeschi}, \citenamefont {Fonseka},
  \citenamefont {Miriametro}, \citenamefont {Tan}, \citenamefont {Jagadish},
  \citenamefont {Capizzi},\ and\ \citenamefont {Polimeni}}]{Zilli2015}%
  \BibitemOpen
  \bibfield  {author} {\bibinfo {author} {\bibfnamefont {A.}~\bibnamefont
  {Zilli}}, \bibinfo {author} {\bibfnamefont {M.~D.}\ \bibnamefont {Luca}},
  \bibinfo {author} {\bibfnamefont {D.}~\bibnamefont {Tedeschi}}, \bibinfo
  {author} {\bibfnamefont {H.~A.}\ \bibnamefont {Fonseka}}, \bibinfo {author}
  {\bibfnamefont {A.}~\bibnamefont {Miriametro}}, \bibinfo {author}
  {\bibfnamefont {H.~H.}\ \bibnamefont {Tan}}, \bibinfo {author} {\bibfnamefont
  {C.}~\bibnamefont {Jagadish}}, \bibinfo {author} {\bibfnamefont
  {M.}~\bibnamefont {Capizzi}}, \ and\ \bibinfo {author} {\bibfnamefont
  {A.}~\bibnamefont {Polimeni}},\ }\href {\doibase 10.1021/acsnano.5b00699}
  {\bibfield  {journal} {\bibinfo  {journal} {ACS Nano}\ }\textbf {\bibinfo
  {volume} {9}},\ \bibinfo {pages} {4277} (\bibinfo {year} {2015})}\BibitemShut
  {NoStop}%
\bibitem [{\citenamefont {Cho}\ \emph {et~al.}(1998)\citenamefont {Cho},
  \citenamefont {Gainer}, \citenamefont {Fischer}, \citenamefont {Song},
  \citenamefont {Keller}, \citenamefont {Mishra},\ and\ \citenamefont
  {DenBaars}}]{Cho1998}%
  \BibitemOpen
  \bibfield  {author} {\bibinfo {author} {\bibfnamefont {Y.-H.}\ \bibnamefont
  {Cho}}, \bibinfo {author} {\bibfnamefont {G.~H.}\ \bibnamefont {Gainer}},
  \bibinfo {author} {\bibfnamefont {A.~J.}\ \bibnamefont {Fischer}}, \bibinfo
  {author} {\bibfnamefont {J.~J.}\ \bibnamefont {Song}}, \bibinfo {author}
  {\bibfnamefont {S.}~\bibnamefont {Keller}}, \bibinfo {author} {\bibfnamefont
  {U.~K.}\ \bibnamefont {Mishra}}, \ and\ \bibinfo {author} {\bibfnamefont
  {S.~P.}\ \bibnamefont {DenBaars}},\ }\href {\doibase
  http://dx.doi.org/10.1063/1.122164} {\bibfield  {journal} {\bibinfo
  {journal} {Appl. Phys. Lett.}\ }\textbf {\bibinfo {volume} {73}},\ \bibinfo
  {pages} {1370} (\bibinfo {year} {1998})}\BibitemShut {NoStop}%
\bibitem [{\citenamefont {Hauswald}\ \emph {et~al.}(2014)\citenamefont
  {Hauswald}, \citenamefont {Corfdir}, \citenamefont {Zettler}, \citenamefont
  {Kaganer}, \citenamefont {Sabelfeld}, \citenamefont
  {Fern{\'{a}}ndez-Garrido}, \citenamefont {Flissikowski}, \citenamefont
  {Consonni}, \citenamefont {Gotschke}, \citenamefont {Grahn}, \citenamefont
  {Geelhaar},\ and\ \citenamefont {Brandt}}]{Hauswald2014}%
  \BibitemOpen
  \bibfield  {author} {\bibinfo {author} {\bibfnamefont {C.}~\bibnamefont
  {Hauswald}}, \bibinfo {author} {\bibfnamefont {P.}~\bibnamefont {Corfdir}},
  \bibinfo {author} {\bibfnamefont {J.~K.}\ \bibnamefont {Zettler}}, \bibinfo
  {author} {\bibfnamefont {V.~M.}\ \bibnamefont {Kaganer}}, \bibinfo {author}
  {\bibfnamefont {K.~K.}\ \bibnamefont {Sabelfeld}}, \bibinfo {author}
  {\bibfnamefont {S.}~\bibnamefont {Fern{\'{a}}ndez-Garrido}}, \bibinfo
  {author} {\bibfnamefont {T.}~\bibnamefont {Flissikowski}}, \bibinfo {author}
  {\bibfnamefont {V.}~\bibnamefont {Consonni}}, \bibinfo {author}
  {\bibfnamefont {T.}~\bibnamefont {Gotschke}}, \bibinfo {author}
  {\bibfnamefont {H.~T.}\ \bibnamefont {Grahn}}, \bibinfo {author}
  {\bibfnamefont {L.}~\bibnamefont {Geelhaar}}, \ and\ \bibinfo {author}
  {\bibfnamefont {O.}~\bibnamefont {Brandt}},\ }\href {\doibase
  10.1103/PhysRevB.90.165304} {\bibfield  {journal} {\bibinfo  {journal} {Phys.
  Rev. B}\ }\textbf {\bibinfo {volume} {90}},\ \bibinfo {pages} {165304}
  (\bibinfo {year} {2014})}\BibitemShut {NoStop}%
\bibitem [{\citenamefont {Titova}\ \emph {et~al.}(2006)\citenamefont {Titova},
  \citenamefont {Hoang}, \citenamefont {Jackson}, \citenamefont {Smith},
  \citenamefont {Yarrison-Rice}, \citenamefont {Kim}, \citenamefont {Joyce},
  \citenamefont {Tan},\ and\ \citenamefont {Jagadish}}]{Titova2006}%
  \BibitemOpen
  \bibfield  {author} {\bibinfo {author} {\bibfnamefont {L.~V.}\ \bibnamefont
  {Titova}}, \bibinfo {author} {\bibfnamefont {T.~B.}\ \bibnamefont {Hoang}},
  \bibinfo {author} {\bibfnamefont {H.~E.}\ \bibnamefont {Jackson}}, \bibinfo
  {author} {\bibfnamefont {L.~M.}\ \bibnamefont {Smith}}, \bibinfo {author}
  {\bibfnamefont {J.~M.}\ \bibnamefont {Yarrison-Rice}}, \bibinfo {author}
  {\bibfnamefont {Y.}~\bibnamefont {Kim}}, \bibinfo {author} {\bibfnamefont
  {H.~J.}\ \bibnamefont {Joyce}}, \bibinfo {author} {\bibfnamefont {H.~H.}\
  \bibnamefont {Tan}}, \ and\ \bibinfo {author} {\bibfnamefont
  {C.}~\bibnamefont {Jagadish}},\ }\href {\doibase 10.1063/1.2364885}
  {\bibfield  {journal} {\bibinfo  {journal} {Appl. Phys. Lett.}\ }\textbf
  {\bibinfo {volume} {89}},\ \bibinfo {pages} {173126} (\bibinfo {year}
  {2006})}\BibitemShut {NoStop}%
\bibitem [{\citenamefont {Rudolph}\ \emph
  {et~al.}(2013{\natexlab{b}})\citenamefont {Rudolph}, \citenamefont
  {Schweickert}, \citenamefont {Mork{\"{o}}tter}, \citenamefont {Hanschke},
  \citenamefont {Hertenberger}, \citenamefont {Bichler}, \citenamefont
  {Koblm{\"{u}}ller}, \citenamefont {Abstreiter},\ and\ \citenamefont
  {Finley}}]{Rudolph2013b}%
  \BibitemOpen
  \bibfield  {author} {\bibinfo {author} {\bibfnamefont {D.}~\bibnamefont
  {Rudolph}}, \bibinfo {author} {\bibfnamefont {L.}~\bibnamefont
  {Schweickert}}, \bibinfo {author} {\bibfnamefont {S.}~\bibnamefont
  {Mork{\"{o}}tter}}, \bibinfo {author} {\bibfnamefont {L.}~\bibnamefont
  {Hanschke}}, \bibinfo {author} {\bibfnamefont {S.}~\bibnamefont
  {Hertenberger}}, \bibinfo {author} {\bibfnamefont {M.}~\bibnamefont
  {Bichler}}, \bibinfo {author} {\bibfnamefont {G.}~\bibnamefont
  {Koblm{\"{u}}ller}}, \bibinfo {author} {\bibfnamefont {G.}~\bibnamefont
  {Abstreiter}}, \ and\ \bibinfo {author} {\bibfnamefont {J.~J.}\ \bibnamefont
  {Finley}},\ }\href {\doibase 10.1088/1367-2630/15/11/113032} {\bibfield
  {journal} {\bibinfo  {journal} {New J. Phys.}\ }\textbf {\bibinfo {volume}
  {15}},\ \bibinfo {pages} {113032} (\bibinfo {year}
  {2013}{\natexlab{b}})}\BibitemShut {NoStop}%
\bibitem [{\citenamefont {Corfdir}\ \emph {et~al.}(2012)\citenamefont
  {Corfdir}, \citenamefont {Dussaigne}, \citenamefont {Teisseyre},
  \citenamefont {Suski}, \citenamefont {Grzegory}, \citenamefont {Lefebvre},
  \citenamefont {Giraud}, \citenamefont {Gani{\`{e}}re}, \citenamefont
  {Grandjean},\ and\ \citenamefont {Deveaud-Pl{\'{e}}dran}}]{Corfdir2012}%
  \BibitemOpen
  \bibfield  {author} {\bibinfo {author} {\bibfnamefont {P.}~\bibnamefont
  {Corfdir}}, \bibinfo {author} {\bibfnamefont {A.}~\bibnamefont {Dussaigne}},
  \bibinfo {author} {\bibfnamefont {H.}~\bibnamefont {Teisseyre}}, \bibinfo
  {author} {\bibfnamefont {T.}~\bibnamefont {Suski}}, \bibinfo {author}
  {\bibfnamefont {I.}~\bibnamefont {Grzegory}}, \bibinfo {author}
  {\bibfnamefont {P.}~\bibnamefont {Lefebvre}}, \bibinfo {author}
  {\bibfnamefont {E.}~\bibnamefont {Giraud}}, \bibinfo {author} {\bibfnamefont
  {J.~D.}\ \bibnamefont {Gani{\`{e}}re}}, \bibinfo {author} {\bibfnamefont
  {N.}~\bibnamefont {Grandjean}}, \ and\ \bibinfo {author} {\bibfnamefont
  {B.}~\bibnamefont {Deveaud-Pl{\'{e}}dran}},\ }\href
  {http://dx.doi.org/10.1063/1.3681816} {\bibfield  {journal} {\bibinfo
  {journal} {J. Appl. Phys.}\ }\textbf {\bibinfo {volume} {111}} (\bibinfo
  {year} {2012})}\BibitemShut {NoStop}%
\bibitem [{\citenamefont {Yu}\ \emph {et~al.}(1996)\citenamefont {Yu},
  \citenamefont {Lycett}, \citenamefont {Roberts},\ and\ \citenamefont
  {Murray}}]{Yu1996}%
  \BibitemOpen
  \bibfield  {author} {\bibinfo {author} {\bibfnamefont {H.}~\bibnamefont
  {Yu}}, \bibinfo {author} {\bibfnamefont {S.}~\bibnamefont {Lycett}}, \bibinfo
  {author} {\bibfnamefont {C.}~\bibnamefont {Roberts}}, \ and\ \bibinfo
  {author} {\bibfnamefont {R.}~\bibnamefont {Murray}},\ }\href {\doibase
  http://dx.doi.org/10.1063/1.117827} {\bibfield  {journal} {\bibinfo
  {journal} {Appl. Phys. Lett.}\ }\textbf {\bibinfo {volume} {69}},\ \bibinfo
  {pages} {4087} (\bibinfo {year} {1996})}\BibitemShut {NoStop}%
\bibitem [{\citenamefont {Suffczy\ifmmode~\acute{n}\else \'{n}\fi{}ski}\ \emph
  {et~al.}(2009)\citenamefont {Suffczy\ifmmode~\acute{n}\else \'{n}\fi{}ski},
  \citenamefont {Dousse}, \citenamefont {Gauthron}, \citenamefont
  {Lema\^{\i}tre}, \citenamefont {Sagnes}, \citenamefont {Lanco}, \citenamefont
  {Bloch}, \citenamefont {Voisin},\ and\ \citenamefont
  {Senellart}}]{Suffczynski2009}%
  \BibitemOpen
  \bibfield  {author} {\bibinfo {author} {\bibfnamefont {J.}~\bibnamefont
  {Suffczy\ifmmode~\acute{n}\else \'{n}\fi{}ski}}, \bibinfo {author}
  {\bibfnamefont {A.}~\bibnamefont {Dousse}}, \bibinfo {author} {\bibfnamefont
  {K.}~\bibnamefont {Gauthron}}, \bibinfo {author} {\bibfnamefont
  {A.}~\bibnamefont {Lema\^{\i}tre}}, \bibinfo {author} {\bibfnamefont
  {I.}~\bibnamefont {Sagnes}}, \bibinfo {author} {\bibfnamefont
  {L.}~\bibnamefont {Lanco}}, \bibinfo {author} {\bibfnamefont
  {J.}~\bibnamefont {Bloch}}, \bibinfo {author} {\bibfnamefont
  {P.}~\bibnamefont {Voisin}}, \ and\ \bibinfo {author} {\bibfnamefont
  {P.}~\bibnamefont {Senellart}},\ }\href {\doibase
  10.1103/PhysRevLett.103.027401} {\bibfield  {journal} {\bibinfo  {journal}
  {Phys. Rev. Lett.}\ }\textbf {\bibinfo {volume} {103}},\ \bibinfo {pages}
  {027401} (\bibinfo {year} {2009})}\BibitemShut {NoStop}%
\bibitem [{\citenamefont {Wolford}\ \emph {et~al.}(1994)\citenamefont
  {Wolford}, \citenamefont {Gilliland}, \citenamefont {Kuech}, \citenamefont
  {Klem}, \citenamefont {Hjalmarson}, \citenamefont {Bradley}, \citenamefont
  {Tsang},\ and\ \citenamefont {Martinsen}}]{Wolford1994}%
  \BibitemOpen
  \bibfield  {author} {\bibinfo {author} {\bibfnamefont {D.~J.}\ \bibnamefont
  {Wolford}}, \bibinfo {author} {\bibfnamefont {G.~D.}\ \bibnamefont
  {Gilliland}}, \bibinfo {author} {\bibfnamefont {T.~F.}\ \bibnamefont
  {Kuech}}, \bibinfo {author} {\bibfnamefont {J.~F.}\ \bibnamefont {Klem}},
  \bibinfo {author} {\bibfnamefont {H.~P.}\ \bibnamefont {Hjalmarson}},
  \bibinfo {author} {\bibfnamefont {J.~A.}\ \bibnamefont {Bradley}}, \bibinfo
  {author} {\bibfnamefont {C.~F.}\ \bibnamefont {Tsang}}, \ and\ \bibinfo
  {author} {\bibfnamefont {J.}~\bibnamefont {Martinsen}},\ }\href {\doibase
  10.1063/1.111901} {\bibfield  {journal} {\bibinfo  {journal} {Appl. Phys.
  Lett.}\ }\textbf {\bibinfo {volume} {64}},\ \bibinfo {pages} {1416} (\bibinfo
  {year} {1994})}\BibitemShut {NoStop}%
\bibitem [{\citenamefont {Demichel}\ \emph {et~al.}(2010)\citenamefont
  {Demichel}, \citenamefont {Heiss}, \citenamefont {Bleuse}, \citenamefont
  {Mariette},\ and\ \citenamefont {{Fontcuberta i Morral}}}]{Demichel2010}%
  \BibitemOpen
  \bibfield  {author} {\bibinfo {author} {\bibfnamefont {O.}~\bibnamefont
  {Demichel}}, \bibinfo {author} {\bibfnamefont {M.}~\bibnamefont {Heiss}},
  \bibinfo {author} {\bibfnamefont {J.}~\bibnamefont {Bleuse}}, \bibinfo
  {author} {\bibfnamefont {H.}~\bibnamefont {Mariette}}, \ and\ \bibinfo
  {author} {\bibfnamefont {A.}~\bibnamefont {{Fontcuberta i Morral}}},\ }\href
  {\doibase 10.1063/1.3519980} {\bibfield  {journal} {\bibinfo  {journal}
  {Appl. Phys. Lett.}\ }\textbf {\bibinfo {volume} {97}},\ \bibinfo {pages}
  {201907} (\bibinfo {year} {2010})}\BibitemShut {NoStop}%
\bibitem [{\citenamefont {Xu}\ and\ \citenamefont {Tang}(1984)}]{Xu1984}%
  \BibitemOpen
  \bibfield  {author} {\bibinfo {author} {\bibfnamefont {Z.~Y.}\ \bibnamefont
  {Xu}}\ and\ \bibinfo {author} {\bibfnamefont {C.~L.}\ \bibnamefont {Tang}},\
  }\href {\doibase http://dx.doi.org/10.1063/1.94880} {\bibfield  {journal}
  {\bibinfo  {journal} {Appl. Phys. Lett.}\ }\textbf {\bibinfo {volume} {44}},\
  \bibinfo {pages} {692} (\bibinfo {year} {1984})}\BibitemShut {NoStop}%
\bibitem [{\citenamefont {Pelouch}\ \emph {et~al.}(1992)\citenamefont
  {Pelouch}, \citenamefont {Ellingson}, \citenamefont {Powers}, \citenamefont
  {Tang}, \citenamefont {Szmyd},\ and\ \citenamefont {Nozik}}]{Pelouch1992}%
  \BibitemOpen
  \bibfield  {author} {\bibinfo {author} {\bibfnamefont {W.~S.}\ \bibnamefont
  {Pelouch}}, \bibinfo {author} {\bibfnamefont {R.~J.}\ \bibnamefont
  {Ellingson}}, \bibinfo {author} {\bibfnamefont {P.~E.}\ \bibnamefont
  {Powers}}, \bibinfo {author} {\bibfnamefont {C.~L.}\ \bibnamefont {Tang}},
  \bibinfo {author} {\bibfnamefont {D.~M.}\ \bibnamefont {Szmyd}}, \ and\
  \bibinfo {author} {\bibfnamefont {A.~J.}\ \bibnamefont {Nozik}},\ }\href
  {\doibase 10.1103/PhysRevB.45.1450} {\bibfield  {journal} {\bibinfo
  {journal} {Phys. Rev. B}\ }\textbf {\bibinfo {volume} {45}},\ \bibinfo
  {pages} {1450} (\bibinfo {year} {1992})}\BibitemShut {NoStop}%
\bibitem [{\citenamefont {Szczytko}\ \emph {et~al.}(2004)\citenamefont
  {Szczytko}, \citenamefont {Kappei}, \citenamefont {Berney}, \citenamefont
  {Morier-Genoud}, \citenamefont {Portella-Oberli},\ and\ \citenamefont
  {Deveaud}}]{Szczytko2004}%
  \BibitemOpen
  \bibfield  {author} {\bibinfo {author} {\bibfnamefont {J.}~\bibnamefont
  {Szczytko}}, \bibinfo {author} {\bibfnamefont {L.}~\bibnamefont {Kappei}},
  \bibinfo {author} {\bibfnamefont {J.}~\bibnamefont {Berney}}, \bibinfo
  {author} {\bibfnamefont {F.}~\bibnamefont {Morier-Genoud}}, \bibinfo {author}
  {\bibfnamefont {M.~T.}\ \bibnamefont {Portella-Oberli}}, \ and\ \bibinfo
  {author} {\bibfnamefont {B.}~\bibnamefont {Deveaud}},\ }\href {\doibase
  10.1103/PhysRevLett.93.137401} {\bibfield  {journal} {\bibinfo  {journal}
  {Phys. Rev. Lett.}\ }\textbf {\bibinfo {volume} {93}},\ \bibinfo {pages}
  {137401} (\bibinfo {year} {2004})}\BibitemShut {NoStop}%
\bibitem [{\citenamefont {Kappei}\ \emph {et~al.}(2005)\citenamefont {Kappei},
  \citenamefont {Szczytko}, \citenamefont {Morier-Genoud},\ and\ \citenamefont
  {Deveaud}}]{Kappei2005}%
  \BibitemOpen
  \bibfield  {author} {\bibinfo {author} {\bibfnamefont {L.}~\bibnamefont
  {Kappei}}, \bibinfo {author} {\bibfnamefont {J.}~\bibnamefont {Szczytko}},
  \bibinfo {author} {\bibfnamefont {F.}~\bibnamefont {Morier-Genoud}}, \ and\
  \bibinfo {author} {\bibfnamefont {B.}~\bibnamefont {Deveaud}},\ }\href
  {\doibase 10.1103/PhysRevLett.94.147403} {\bibfield  {journal} {\bibinfo
  {journal} {Phys. Rev. Lett.}\ }\textbf {\bibinfo {volume} {94}},\ \bibinfo
  {pages} {147403} (\bibinfo {year} {2005})}\BibitemShut {NoStop}%
\bibitem [{\citenamefont {Amo}\ \emph {et~al.}(2007)\citenamefont {Amo},
  \citenamefont {Martín}, \citenamefont {Viña}, \citenamefont {Toropov},\
  and\ \citenamefont {Zhuravlev}}]{Amo2007}%
  \BibitemOpen
  \bibfield  {author} {\bibinfo {author} {\bibfnamefont {A.}~\bibnamefont
  {Amo}}, \bibinfo {author} {\bibfnamefont {M.~D.}\ \bibnamefont {Martín}},
  \bibinfo {author} {\bibfnamefont {L.}~\bibnamefont {Viña}}, \bibinfo
  {author} {\bibfnamefont {A.~I.}\ \bibnamefont {Toropov}}, \ and\ \bibinfo
  {author} {\bibfnamefont {K.~S.}\ \bibnamefont {Zhuravlev}},\ }\href {\doibase
  10.1063/1.2722786} {\bibfield  {journal} {\bibinfo  {journal} {Journal of
  Applied Physics}\ }\textbf {\bibinfo {volume} {101}},\ \bibinfo {pages}
  {081717} (\bibinfo {year} {2007})}\BibitemShut {NoStop}%
\bibitem [{Note1()}]{Note1}%
  \BibitemOpen
  \bibinfo {note} {The measured carrier densities are significantly smaller
  than those obtained from simple estimates based on the energy fluence per
  pulse. These differences arise from the reflection of the laser from the
  surface, surface and interface recombination, diffusion and plasma
  expansion.}\BibitemShut {Stop}%
\bibitem [{\citenamefont {Heiss}\ \emph {et~al.}(2014)\citenamefont {Heiss},
  \citenamefont {Russo-Averchi}, \citenamefont {Dalmau-Mallorqu{\'{\i}}},
  \citenamefont {T{\"{u}}t{\"{u}}nc{\"{u}}oğlu}, \citenamefont {Matteini},
  \citenamefont {R{\"{u}}ffer}, \citenamefont {Conesa-Boj}, \citenamefont
  {Demichel}, \citenamefont {Alarcon-Llad{\'{o}}},\ and\ \citenamefont
  {{Fontcuberta i Morral}}}]{Heiss2014}%
  \BibitemOpen
  \bibfield  {author} {\bibinfo {author} {\bibfnamefont {M.}~\bibnamefont
  {Heiss}}, \bibinfo {author} {\bibfnamefont {E.}~\bibnamefont
  {Russo-Averchi}}, \bibinfo {author} {\bibfnamefont {A.}~\bibnamefont
  {Dalmau-Mallorqu{\'{\i}}}}, \bibinfo {author} {\bibfnamefont
  {G.}~\bibnamefont {T{\"{u}}t{\"{u}}nc{\"{u}}oğlu}}, \bibinfo {author}
  {\bibfnamefont {F.}~\bibnamefont {Matteini}}, \bibinfo {author}
  {\bibfnamefont {D.}~\bibnamefont {R{\"{u}}ffer}}, \bibinfo {author}
  {\bibfnamefont {S.}~\bibnamefont {Conesa-Boj}}, \bibinfo {author}
  {\bibfnamefont {O.}~\bibnamefont {Demichel}}, \bibinfo {author}
  {\bibfnamefont {E.}~\bibnamefont {Alarcon-Llad{\'{o}}}}, \ and\ \bibinfo
  {author} {\bibfnamefont {A.}~\bibnamefont {{Fontcuberta i Morral}}},\ }\href
  {\doibase 10.1088/0957-4484/25/1/014015} {\bibfield  {journal} {\bibinfo
  {journal} {Nanotechnology}\ }\textbf {\bibinfo {volume} {25}},\ \bibinfo
  {pages} {014015} (\bibinfo {year} {2014})}\BibitemShut {NoStop}%
\bibitem [{\citenamefont {Tedeschi}\ \emph {et~al.}(2016)\citenamefont
  {Tedeschi}, \citenamefont {De~Luca}, \citenamefont {Fonseka}, \citenamefont
  {Gao}, \citenamefont {Mura}, \citenamefont {Tan}, \citenamefont {Rubini},
  \citenamefont {Martelli}, \citenamefont {Jagadish}, \citenamefont {Capizzi},\
  and\ \citenamefont {Polimeni}}]{Tedeschi2016}%
  \BibitemOpen
  \bibfield  {author} {\bibinfo {author} {\bibfnamefont {D.}~\bibnamefont
  {Tedeschi}}, \bibinfo {author} {\bibfnamefont {M.}~\bibnamefont {De~Luca}},
  \bibinfo {author} {\bibfnamefont {H.~A.}\ \bibnamefont {Fonseka}}, \bibinfo
  {author} {\bibfnamefont {Q.}~\bibnamefont {Gao}}, \bibinfo {author}
  {\bibfnamefont {F.}~\bibnamefont {Mura}}, \bibinfo {author} {\bibfnamefont
  {H.~H.}\ \bibnamefont {Tan}}, \bibinfo {author} {\bibfnamefont
  {S.}~\bibnamefont {Rubini}}, \bibinfo {author} {\bibfnamefont
  {F.}~\bibnamefont {Martelli}}, \bibinfo {author} {\bibfnamefont
  {C.}~\bibnamefont {Jagadish}}, \bibinfo {author} {\bibfnamefont
  {M.}~\bibnamefont {Capizzi}}, \ and\ \bibinfo {author} {\bibfnamefont
  {A.}~\bibnamefont {Polimeni}},\ }\href {\doibase
  10.1021/acs.nanolett.6b00251} {\bibfield  {journal} {\bibinfo  {journal}
  {Nano Lett.}\ }\textbf {\bibinfo {volume} {16}},\ \bibinfo {pages} {3085}
  (\bibinfo {year} {2016})}\BibitemShut {NoStop}%
\bibitem [{\citenamefont {Gourdon}\ and\ \citenamefont
  {Lavallard}(1989)}]{Gourdon1989}%
  \BibitemOpen
  \bibfield  {author} {\bibinfo {author} {\bibfnamefont {C.}~\bibnamefont
  {Gourdon}}\ and\ \bibinfo {author} {\bibfnamefont {P.}~\bibnamefont
  {Lavallard}},\ }\href {\doibase 10.1002/pssb.2221530222} {\bibfield
  {journal} {\bibinfo  {journal} {Phys. Status Solidi B}\ }\textbf {\bibinfo
  {volume} {153}},\ \bibinfo {pages} {641} (\bibinfo {year}
  {1989})}\BibitemShut {NoStop}%
\bibitem [{\citenamefont {Colombo}\ \emph {et~al.}(2009)\citenamefont
  {Colombo}, \citenamefont {Hei{\ss}}, \citenamefont {Gr{\"{a}}tzel},\ and\
  \citenamefont {{Fontcuberta i Morral}}}]{Colombo2009}%
  \BibitemOpen
  \bibfield  {author} {\bibinfo {author} {\bibfnamefont {C.}~\bibnamefont
  {Colombo}}, \bibinfo {author} {\bibfnamefont {M.}~\bibnamefont {Hei{\ss}}},
  \bibinfo {author} {\bibfnamefont {M.}~\bibnamefont {Gr{\"{a}}tzel}}, \ and\
  \bibinfo {author} {\bibfnamefont {A.}~\bibnamefont {{Fontcuberta i
  Morral}}},\ }\href {\doibase 10.1063/1.3125435} {\bibfield  {journal}
  {\bibinfo  {journal} {Appl. Phys. Lett.}\ }\textbf {\bibinfo {volume} {94}},\
  \bibinfo {pages} {173108} (\bibinfo {year} {2009})}\BibitemShut {NoStop}%
\bibitem [{\citenamefont {Dimakis}\ \emph {et~al.}(2014)\citenamefont
  {Dimakis}, \citenamefont {Jahn}, \citenamefont {Ramsteiner}, \citenamefont
  {Tahraoui}, \citenamefont {Grandal}, \citenamefont {Kong}, \citenamefont
  {Marquardt}, \citenamefont {Trampert}, \citenamefont {Riechert},\ and\
  \citenamefont {Geelhaar}}]{Dimakis2014}%
  \BibitemOpen
  \bibfield  {author} {\bibinfo {author} {\bibfnamefont {E.}~\bibnamefont
  {Dimakis}}, \bibinfo {author} {\bibfnamefont {U.}~\bibnamefont {Jahn}},
  \bibinfo {author} {\bibfnamefont {M.}~\bibnamefont {Ramsteiner}}, \bibinfo
  {author} {\bibfnamefont {A.}~\bibnamefont {Tahraoui}}, \bibinfo {author}
  {\bibfnamefont {J.}~\bibnamefont {Grandal}}, \bibinfo {author} {\bibfnamefont
  {X.}~\bibnamefont {Kong}}, \bibinfo {author} {\bibfnamefont {O.}~\bibnamefont
  {Marquardt}}, \bibinfo {author} {\bibfnamefont {A.}~\bibnamefont {Trampert}},
  \bibinfo {author} {\bibfnamefont {H.}~\bibnamefont {Riechert}}, \ and\
  \bibinfo {author} {\bibfnamefont {L.}~\bibnamefont {Geelhaar}},\ }\href
  {\doibase 10.1021/nl500428v} {\bibfield  {journal} {\bibinfo  {journal} {Nano
  Lett.}\ }\textbf {\bibinfo {volume} {14}},\ \bibinfo {pages} {2604} (\bibinfo
  {year} {2014})}\BibitemShut {NoStop}%
\bibitem [{\citenamefont {Funk}\ \emph {et~al.}(2013)\citenamefont {Funk},
  \citenamefont {Royo}, \citenamefont {Zardo}, \citenamefont {Rudolph},
  \citenamefont {Morkötter}, \citenamefont {Mayer}, \citenamefont {Becker},
  \citenamefont {Bechtold}, \citenamefont {Matich}, \citenamefont {Döblinger},
  \citenamefont {Bichler}, \citenamefont {Koblmüller}, \citenamefont {Finley},
  \citenamefont {Bertoni}, \citenamefont {Goldoni},\ and\ \citenamefont
  {Abstreiter}}]{Funk2013}%
  \BibitemOpen
  \bibfield  {author} {\bibinfo {author} {\bibfnamefont {S.}~\bibnamefont
  {Funk}}, \bibinfo {author} {\bibfnamefont {M.}~\bibnamefont {Royo}}, \bibinfo
  {author} {\bibfnamefont {I.}~\bibnamefont {Zardo}}, \bibinfo {author}
  {\bibfnamefont {D.}~\bibnamefont {Rudolph}}, \bibinfo {author} {\bibfnamefont
  {S.}~\bibnamefont {Morkötter}}, \bibinfo {author} {\bibfnamefont
  {B.}~\bibnamefont {Mayer}}, \bibinfo {author} {\bibfnamefont
  {J.}~\bibnamefont {Becker}}, \bibinfo {author} {\bibfnamefont
  {A.}~\bibnamefont {Bechtold}}, \bibinfo {author} {\bibfnamefont
  {S.}~\bibnamefont {Matich}}, \bibinfo {author} {\bibfnamefont
  {M.}~\bibnamefont {Döblinger}}, \bibinfo {author} {\bibfnamefont
  {M.}~\bibnamefont {Bichler}}, \bibinfo {author} {\bibfnamefont
  {G.}~\bibnamefont {Koblmüller}}, \bibinfo {author} {\bibfnamefont {J.~J.}\
  \bibnamefont {Finley}}, \bibinfo {author} {\bibfnamefont {A.}~\bibnamefont
  {Bertoni}}, \bibinfo {author} {\bibfnamefont {G.}~\bibnamefont {Goldoni}}, \
  and\ \bibinfo {author} {\bibfnamefont {G.}~\bibnamefont {Abstreiter}},\
  }\href {\doibase 10.1021/nl403561w} {\bibfield  {journal} {\bibinfo
  {journal} {Nano Lett.}\ }\textbf {\bibinfo {volume} {13}},\ \bibinfo {pages}
  {6189} (\bibinfo {year} {2013})}\BibitemShut {NoStop}%
\bibitem [{\citenamefont {Royo}\ \emph {et~al.}(2015)\citenamefont {Royo},
  \citenamefont {Segarra}, \citenamefont {Bertoni}, \citenamefont {Goldoni},\
  and\ \citenamefont {Planelles}}]{Royo2015}%
  \BibitemOpen
  \bibfield  {author} {\bibinfo {author} {\bibfnamefont {M.}~\bibnamefont
  {Royo}}, \bibinfo {author} {\bibfnamefont {C.}~\bibnamefont {Segarra}},
  \bibinfo {author} {\bibfnamefont {A.}~\bibnamefont {Bertoni}}, \bibinfo
  {author} {\bibfnamefont {G.}~\bibnamefont {Goldoni}}, \ and\ \bibinfo
  {author} {\bibfnamefont {J.}~\bibnamefont {Planelles}},\ }\href {\doibase
  10.1103/PhysRevB.91.115440} {\bibfield  {journal} {\bibinfo  {journal} {Phys.
  Rev. B}\ }\textbf {\bibinfo {volume} {91}},\ \bibinfo {pages} {115440}
  (\bibinfo {year} {2015})}\BibitemShut {NoStop}%
\end{thebibliography}%

\end{document}